\documentclass[two column]{aastex701}

\graphicspath{{./}{figures/}}

\usepackage{hyperref}

\shortauthors{liu et al.}
\newcommand{\emailaddress}{liuyuhua@shao.ac.cn}

\newcommand{\dotarcsec}{\rlap{.}\arcsec}
\usepackage{csquotes}
\usepackage{amsmath}
\usepackage{mathtools}
\usepackage[inline]{enumitem} 
\usepackage{float}
\graphicspath{{./}{figures/}}
\usepackage{filecontents}
\usepackage{graphicx}
\usepackage{booktabs}
\usepackage{CJKutf8}

\begin{document}

\title{The Dense Gas Structures Around MMS\,2/OMC-3 Traced by C$^{18}$O Emission}

\author[0009-0009-2263-5502]{yuhua liu \begin{CJK}{UTF8}{gbsn}(柳玉华)\end{CJK}}
\affiliation{Shanghai Astronomical Observatory, Chinese Academy of Sciences, 80 Nandan Road, Shanghai 200030, P.R.China; \href{mailto:\emailaddress}{\emailaddress}}
\email{liuyuhua@shao.ac.cn}

\author[0000-0002-7287-4343]{satoko takahashi \begin{CJK}{UTF8}{ipxm}(髙橋智子)\end{CJK}}%{min}
\affiliation{National Astronomical Observatory of Japan, 2-21-1 Osawa, Mitaka, Tokyo 181-8588, Japan}
\affiliation{Department of Astronomical Science, School of Physical Sciences, The Graduate University for Advanced Studies, SOKENDAI, 2-21-1 Osawa, Mitaka, Tokyo 181-8588, Japan}
\email{satoko.takahashi@nao.ac.jp}

\author[0000-0002-0963-0872]{Masahiro N. Machida}
\affiliation{Department of Earth and Planetary Sciences, Faculty of Science, Kyushu University, Fukuoka 819-0395, Japan}
\email{machida.masahiro.018@m.kyushu-u.ac.jp}

\begin{abstract}
We report the Atacama Large Millimeter/submillimeter Array observations of 1.3\,mm continuum, C$^{18}$O\,($J=2$$-$$1$), and N$_{2}$D$^{+}$\,($J=3$$-$$2$) toward the millimeter multi-system MMS\,2 in the Orion Molecular Cloud-3 region, at an angular resolution of 1\dotarcsec59 ($\sim$620 au). MMS\,2 is in the flat-spectrum phase, with its protostars in the late stage of the main accretion phase. We detect the centrally condensed structures traced by C$^{18}$O emission associated with the 1.3\,mm continuum sources MMS\,2-North and MMS\,2-South, which are spatially resolved and unresolved, respectively. The estimated diameter and mass of the centrally condensed structure associated with MMS\,2-North are 1\dotarcsec64 ($\sim$640 au) and 3.0$\times$10$^{-4}$\textit{M}$_{\odot}$, respectively. We also detect an extended structure traced by C$^{18}$O emission associated the circumbinary envelope at scale of 5\dotarcsec70 ($\sim$2240 au) with an estimated gas mass of 1.4$\times$10$^{-2}$\,$M_{\odot}$. In addition, we detect a filamentary structure traced by N$_2$D$^+$ emission to the south of MMS\,2, which is spatially offset from the C$^{18}$O emission. This offset may be attributed to the influence of the warm surrounding environment.
\end{abstract}

 %Circumstellar disks (235)
\section{Introduction}
%Circumstellar disks surrounding the young stellar objects consist of material from star formation and also serve as nurseries for planetary systems.materials surround the protostars rapidly rotates especially in the inner region, establishing the Keplerian motion of the disk. 

%Recently, such substructures were also observed toward some Class I disks \citep[e.g.,][]{n.ohashi2023,maureira2024,shoshi2024}. This suggests that the planet formation might begin as early as Class I phase.
Stars form through the gravitational collapse of dense cores within molecular clouds \citep[e.g.,][]{shu1987}. Stars evolve over different phases, with those in the earliest stages of formation known as young stellar objects (YSOs). The YSOs, such as Class 0, Class I, Class II, and flat‑spectrum sources, are classified based on their spectral energy distributions (SEDs). In the early Class 0/I evolutionary stages, the infalling envelope acts as the mass reservoir. The angular momentum of the infalling gas produces a circumstellar disk around the protostar \citep[e.g.,][]{terebey1984}. The molecular line observations have further demonstrated that the kinematic structures of some circumstellar disks are consistent with Keplerian rotation, and such Keplerian disks have been observed around Class 0/I sources \citep[e.g.,][]{hara2013,murillo2013,ohashi2014,yen2014,matsushita2019very,sai2020,kido2023,n.ohashi2023}. In the late Class II (T-Tauri) phase, the YSOs are ubiquitously surrounded by protoplanetary disks showing Keplerian rotation \citep[e.g.,][]{dutrey1994,handa1995,dutrey1998,simon2000,isella2007,pietu2007,teague2022}.

%During the collapse of the core, the initial angular momentum and magnetic field form a structure known as protostellar infalling envelope \citep[e.g.,][]{shu1987,momose1998,ohashi2014,oya2016,sakai2016}.
%the embedded Class 0/I protostellar sources \textbf{typically} have \textbf{a} positive spectral index\footnote{Spectral index ($\alpha$) describes an exponential factor relating the flux density of a radio source to its frequency, which is used to classify YSOs into the evolutionary stages in the near- and mid-infrared regions, given by $\alpha=\frac{d\textnormal{log}(\lambda F_{\lambda})}{d(\textnormal{log($\lambda$)})}$, where $\lambda$ is the wavelength and $F_{\lambda}$ is the flux density \citep[e.g.,][]{lada1987,andre1994,greene1994}.}, the transitional flat-spectrum sources have \textbf{indices} between 0.3 and $-$\,0.3, and the Class II sources have negative \textbf{indices} between $-$\,0.3 and $-$\,1.6 \citep{furlan2016herschel}. 

%There are numerous of these disks exhibit sub-structures such as rings, gaps, and spirals, providing observational evidence of ongoing planet formation \citep[e.g.,][]{andrews2018,shoshi2025}. 

The early-stage Class 0/I circumstellar disks remain embedded within the envelopes while undergoing active accretion. However, the mass accretion rates gradually decrease over time \citep{andre2000,fischer2017} as the envelopes dissipate \citep{jorgensen2009}. Subsequently, the late-stage Class II protoplanetary disks show the significantly reduced accretion rates compared to those of the early-stage \citep{fiorellino2021,fiorellino2023}, and their surrounding envelopes have mostly dissipated. The flat-spectrum phase that bridges between the Class 0/I and Class II phases is considered as the latest evolutionary stage of the main accretion phase. Studying the transitional flat-spectrum sources provides critical insights into the mechanisms governing the final stage of the main accretion phase, including the dense gas dissipation, the evolving structure, and the dynamics of the disk. Moreover, substructures such as rings, gaps, and spirals considered to be associated with ongoing planet formation observed toward Class II protoplanetary disks \citep[e.g.,][]{andrews2018}. They have also been identified toward the disks around Class 0/I/flat-spectrum sources \citep[e.g.,][]{n.ohashi2023,shoshi2025}. These suggest that the planet formation may begin during the main accretion phase. Therefore, understanding the mechanisms governing the final stage of the main accretion phase is also essential for determining the initial conditions of planet formation.

%\citep{chini1997dust,nielbock2003stellar,takahashi2013hierarchical,furlan2016herschel,tobin2020vla}.
Orion Molecular Cloud-3 (OMC-3) is located in the nearest giant molecular cloud, the Orion-A Giant Molecular Cloud at a distance of $D=393$ pc \citep{tobin2020vla}. The characteristics of several sources at the early evolutionary stages have already been reported in this region \citep[e.g.,][]{takahashi2006millimeter,takahashi2009,takahashi2012spatially,takahashi2012molecular,takahashi2013hierarchical, hull2014, matsushita2019very, takahashi2019alma, liu2021, morii2021revealing, zielinski2022,hirano2024, hsu2024,takahashi2024,huang2024}. For instance, the Class 0 source MMS\,3 has shown the detection of a flatten envelope traced by C$^{18}$O\,($J=2$$-$$1$) emission \citep{morii2021revealing}, while another Class 0 source MMS\,5 has detected a condensed structured in C$^{18}$O\,($J=2$$-$$1$) emission showing the rotational motion consistent with both Keplerian rotation and angular momentum conservation \citep{matsushita2019very}. In addition, the CO\,($J=2$$-$$1$) and SiO\,($J=5$$-$$4$) observations revealed collimated bipolar outflows and jets associated with Class 0 sources MMS\,1 \citep{takahashi2024} and MMS\,5 \citep{matsushita2019very}, and the CO\,($J=3$$-$$2$) and HCN\,($J=4$$-$$3$) revealed another Class 0 source MMS\,6 driving an extremely compact bipolar outflow with a one-sided length of $\sim$1000\,au \citep{takahashi2012molecular}. A disk‑like dense gas envelope exhibiting dispersing motions, traced by H$^{13}$CO$^{+}$\,($J=1$$-$$0$) emission, has been reported toward the Class I source MMS\,7 \citep{takahashi2006millimeter}. However, the dense gas structure of the more evolved flat‑spectrum source MMS\,2 in this region has not yet been investigated.

%the high-density core associated with the prestellar source MMS\,4 has been reported as a candidate of first hydrostatic core \citep{hirano2024, liu2024}. 

MMS\,2 is the flat-spectrum source, which is also one of the most evolved sources in the OMC-3 region \citep{furlan2016herschel,tobin2020vla}. MMS\,2 was first identified by \cite{chini1997dust} in the 1300\,$\mu$m survey, also known as CSO\,6 \citep{lis1998350} and SMM\,3 \citep{takahashi2013hierarchical}. The 10\,$\mu$m observations found that this object is a binary system consisting of two infrared sources showing the Class I-type spectral index \citep{nielbock2003stellar}. The separation between these two sources is $1\dotarcsec3$ \citep{nielbock2003stellar}. MMS\,2 was also identified as HOPS-92 by \cite{furlan2016herschel} through the multi-wavelength infrared observations. MMS\,2 was subsequently classified as a flat-spectrum source based on the SED study \citep{furlan2016herschel}. The bolometric luminosity of this source is estimated to be $\sim$17.6\,$L_{\odot}$ \citep{furlan2016herschel}. In the recent Atacama Large Millimeter/submillimeter Array (ALMA) 0.87\,mm observations, HOPS-92 was further resolved into three sources, HOPS-92-A-A, HOPS-92-A-B, and HOPS-92-B \citep{tobin2020vla}. Subsequently, the ALMA 1.1\,mm continuum observations have also resolved this object into a triple protostellar system with MMS\,2-North-A, MMS\,2-North-B, and MMS\,2-South \citep{liu2024}, which are associated with HOPS-92-A-A, HOPS-92-A-B, and HOPS-92-B, respectively. The separation between MMS\,2-North-A (Northern dominant component) and MMS\,2-South is $\sim$1$\dotarcsec$4 \citep{liu2024}, which is consistent with the separation between HOPS-92-A-A and HOPS-92-B \citep{tobin2020vla}. 
%in which its mid-infrared (4.5$-$24 $\mu$m) spectral index is between $-\,0.3$ and 0.3 

The ALMA 1.1 mm polarization observations revealed several features consistent with dust self-scattering, including low mean polarization fractions of $\sim$1\%, azimuthal polarization vectors in the nearly face-on disk of MMS\,2-North-A, polarization vectors predominantly aligned with the minor axis of the inclined disk of MMS\,2-South, and a clear offset of the polarization intensity peak toward the near side of the MMS\,2-South disk \citep{liu2024}. These results support self-scattering as a possible origin of the observed polarization in both sources and the detection of self-scattering further suggests the presence of dust grain growth \citep{liu2024}. %The ALMA 1.1\,mm polarization observations revealed that the polarization vectors in MMS\,2-North-A were distributed in an azimuthal direction, while the vectors in MMS\,2-South were mainly aligned with minor axis of the disk. The mean polarization fractions for both sources are as low as $\sim$1\%  \citep{liu2024}. Furthermore, a clear peak positional shift of polarization intensity toward the nearside of the disk was identified from MMS\,2-South. All these features support self-scattering as one of the possible origins of the polarization observed toward these two sources  \citep{liu2024}. The polarization features of MMS\,2-North-A are consistent with those expected from self-scattering in a face-on disk, while the features of MMS\,2-South align with those due to self-scattering in an inclined disk \citep{liu2024}. The detection of polarization due to self-scattering also suggests that existence of the dust grains growth \citep{liu2024}. 
As for the molecular line observations toward this object, a bipolar CO outflow with one-sided length of $\sim$0.2 pc elongated along the east-west direction, was observed in previous studies \citep{aso2000dense,williams2003high,takahashi2008millimeter,tanabe2019nobeyama}. Recent CARMA-Nobeyama Radio Observatory Orion (CARMA-NRO Orion) survey reported a more elongated one-sided length of $\sim$0.48 pc associated with this outflow \citep{feddersen2020carma}. In this work, we study the physical properties and the dynamics of MMS\,2, and also report the gas structure associated with MMS\,2 traced by C$^{18}$O emission for the first time.
%Furthermore, the Infrared Imager (IRIM) by \cite{yu1997shock} detected an east-west oriented outflow consisting of a chain of H$_2$ knots associated with MMS\,2. 

We describe the ALMA observations, data reduction, and imaging in Section\,\ref{sec:obser}. The results are summarized in Section\,\ref{sec:results}. We discuss the kinematics of the dense gas structure, dense gas properties, and the evolutionary status in Section\,\ref{sec:discussion}. Finally, we present the conclusions in Section\,\ref{sec:conclusion}.

%Molecular lines such as C$^{18}$O are used to trace the and characterize the disk \citep[e.g.,][]{trapman2022}.
\section{Observations, Data Reduction and Imaging} \label{sec:obser}

\subsection{Observations and Data Reduction}
The ALMA 1.3\,mm (Band 6) continuum observations were made in Cycle 3 (2015.1.00341.S; P.I. S. Takahashi), 2016 January 29 (low angular resolution data), and 2016 September 18 and 19 (high angular resolution data). The phase center was set toward the millimeter source MMS\,2 at R.A. (J2000) = 5$^{\textnormal{h}}$35$^{\textnormal{m}}$18$\rlap{.}^{\textnormal{s}}$30, decl.(J2000) = $-$05$^{\circ}$00$^{\prime}$33$\dotarcsec$01. Forty-eight and around forty of the 12-m antennas were operated for the low angular resolution and high angular resolution observations, respectively. The total on-source time were $\sim$4 minutes (low angular resolution data) and $\sim$16 minutes (high angular resolution data). The full width at half maximum (FWHM) of the ALMA primary beam is $\sim$27$\arcsec$ for both data sets. The projected baselines range between 8.6 and 238 k$\lambda$ (low angular resolution data) and between 8.9 and 2418 k$\lambda$ (high angular resolution data). The corresponding maximum recoverable sizes are $\sim$9$\dotarcsec$8 and $\sim$2$\dotarcsec$0, respectively. Four spectral lines including CO~($J$ = 2$-$1; 230.538 GHz), N$_{2}$D$^{+}$~($J$ = 3$-$2; 231.322 GHz), SiO~($J$ = 5$-$4; 217.105 GHz), and C$^{18}$O~($J$ = 2$-$1; 219.560 GHz) were observed at velocity resolutions of 0.37\,km\,s$^{-1}$,
0.046\,km\,s$^{-1}$, 0.39\,km\,s$^{-1}$, and 0.048\,km\,s$^{-1}$,  respectively. Line-free channels corresponding to the effective bandwidths of 823\,MHz (low angular resolution data) and 839\,MHz (high angular resolution data) were allocated for imaging the 1.3\,mm continuum emissions. The observational parameters are listed in Table~\ref{table:parameters}. The data reduction were performed with the Common Astronomy Software Applications \citep[CASA;][]{bean2022casa} package version 4.6.0 (low angular resolution data) and version 4.7.0 (high angular resolution data).

\subsection{Imaging}
The imaging was performed with the CASA \citep{bean2022casa} package version 6.6.3. The CLEANed images of both 1.3\,mm continuum and molecular line emissions were made using a CASA task ``\texttt{tclean}'' with Briggs weighting (robust = 0.5) for both the low and high angular resolution data sets. For the line emission, the velocity width of 0.1 km\,s$^{-1}$ was used for imaging the molecular lines N$_{2}$D$^{+}$ and C$^{18}$O, and the velocity width of 0.5 km\,s$^{-1}$ was used for imaging the molecular line SiO. The N$_{2}$D$^{+}$ and C$^{18}$O emissions are detected at the $\ge$3$\sigma$ level only in the low angular resolution data set, while these two spectral lines are not detected in the high angular resolution data set. The SiO emission is not detected in either the low or high angular resolution data set. The CO data were already reported in our ALMA 1.1\,mm polarization study \citep{liu2024}. 

In this paper, we present images made from the low angular resolution data set, in which the 1.3\,mm continuum, C$^{18}$O, and N$_{2}$D$^{+}$ emissions are detected at the $\ge$3$\sigma$ level. The moment 0, 1, and 2 maps of the molecular line emissions were made using the CASA task ``\texttt{immoment}'', with moment 1 and 2 maps were produced using a 3$\sigma$ threshold. For both the moment 0 and channel maps, we note the presence of negative features at the $\sim$3$-$6$\sigma$ level, likely arising from sidelobe artifacts. The intensities of these negative components are $\lesssim20\%$ of the peak intensity. Due to the inherent complexity of deconvolution in interferometric imaging, completely removing such artifacts is technically challenging. Nevertheless, these features are minor and do not affect the main scientific conclusions. The position-velocity (PV) diagram was made using the CASA task ``\texttt{impv}''. Further details of the PV diagram are provided in Section\,\ref{4.1}.

\begin{figure*}[ht!]
%\vspace{-25\baselineskip}
\gridline{\hspace{-1.5\baselineskip}
          \fig{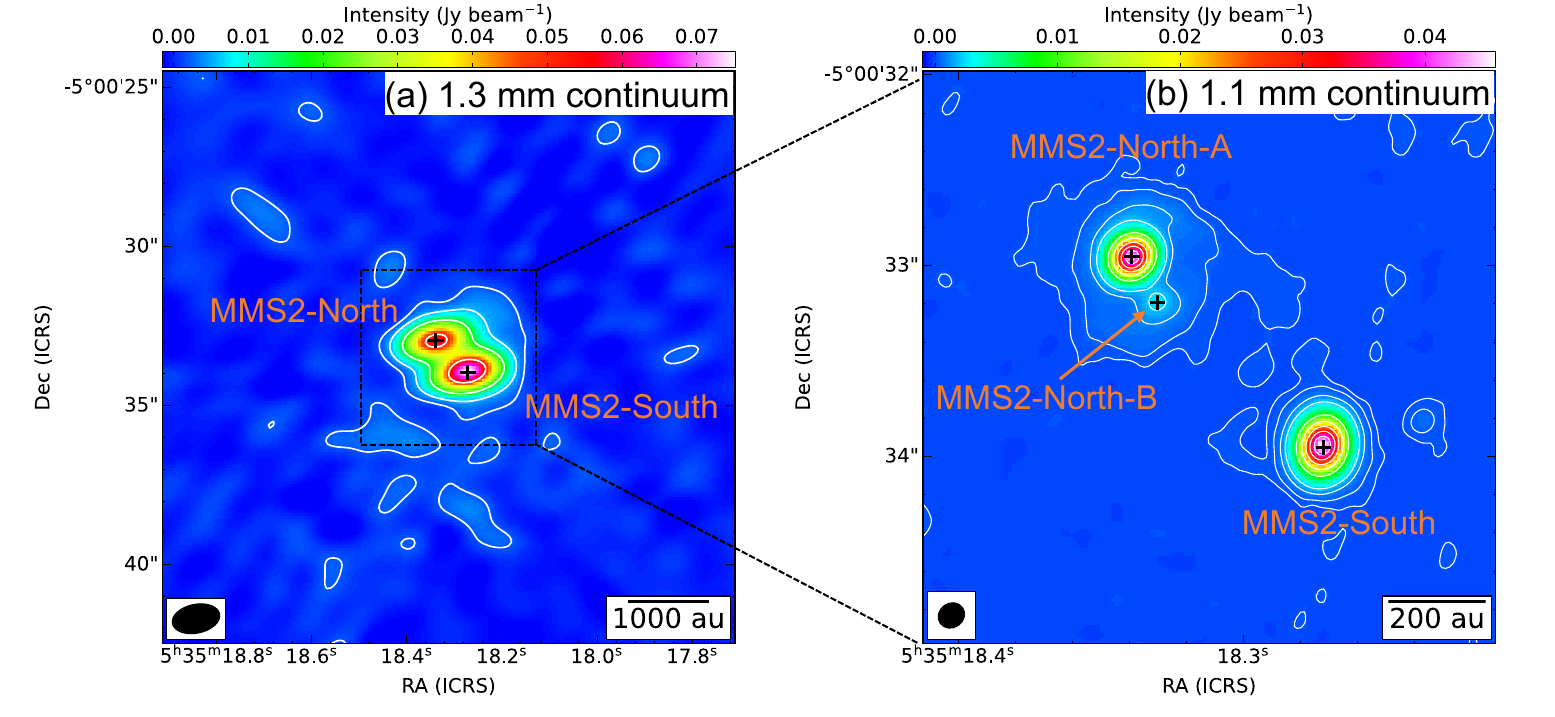}{1\textwidth}{}
          }
%\gridline{\hspace{-1.5\baselineskip}\fig{lar-co.pdf}{0.5\textwidth}{}}
\vspace{-1\baselineskip}
\caption{
    The color and white contours show the continuum emission obtained from (a) ALMA 1.3\,mm low angular resolution data, and (b) ALMA 1.1\,mm high angular resolution data \citep{liu2024}. The white contour levels are [3, 9, 27, 81] $\times$ $\sigma$ (1$\sigma$ = 0.61 mJy\,beam$^{-1}$) for panel (a), and [3, 9, 27, 60, 260, 460, 660, 860, 1060] $\times$ $\sigma$ (1$\sigma$ = 0.04 mJy\,beam$^{-1}$) for panel (b). The black crosses indicate the peak positions of MMS\,2-North and MMS\,2-South, respectively in panel (a), and the peak positions of MMS\,2-North-A, MMS\,2-North-B, and MMS\,2-South, respectively in panel (b). The synthesized beam size is denoted by a filled black ellipse in the bottom left corner for each panel. 
    }
\label{fig-cont}         
\end{figure*}

To compare with the continuum emission derived from the 1.3\,mm low angular resolution data, we use the continuum image (Figure\,\ref{fig-cont}\,(b)) obtained from our previous 1.1\,mm polarization study \citep{liu2024}. This image has an angular resolution of 0\dotarcsec14, which approximately 10 times higher than that of the 1.3\,mm low angular resolution image of 1\dotarcsec52. Although the 1.3\,mm continuum emission is also detected in the high angular resolution configuration, the 1.1\,mm continuum image \citep{liu2024} provides a better signal-to-noise ratio (S/N) and overall image quality. The 1.1\,mm continuum image has an angular resolution $\sim$1.5 times higher than that of the high angular resolution 1.3\,mm continuum image, with a S/N ratio $\sim$1.2 times higher. For these reasons, we use the 1.1\,mm continuum image \citep{liu2024} for comparison in Section\,\ref{result3.1}. Table~\ref{tab:line statistics} summarizes the rms noise level, synthesized beam size, and velocity width for the images present in this paper.

%\textbf{In this paper, the term ``velocity resolution" refers to the instrumental resolution, and the term ``velocity width" refers to the post-processing of imaging.}
%We did not use the 1.3\,mm high angular resolution continuum image because the resolution and signal-to-noise ratio (S/N) of the 1.1\,mm polarized continuum image are better than that of the 1.3\,mm high angular resolution continuum image.
%C$^{18}$O~($J$ = 2$-$1) 
%Some of the molecular line data were already presented in our previous studies \citep{takahashi2019alma,matsushita2019very,morii2021revealing,takahashi2024,liu2024}.}

\begin{table*}[ht!]
%\label{parameters}
{\scriptsize
\begin{center}
\caption{\small ALMA Observations}
\label{table:parameters}
\begin{tabular}{ccc}
\hline\hline \noalign {\smallskip}
%Parameters & X368f, X435c, \& X4b9d & X336b \& X3d1d & X777f \& X8048\\
Parameters &  Low angular resolution  & High angular resolution  \\
\hline
%Observed sources & \multicolumn{2}{c}{MMS\,2}   & MMS\,3 and MMS\,4  & MMS\,1$-$7\\
Observing date (YYYY-MM-DD)	& 2016-01-29	&  2016-09-18 and -19\\
Number of antennas		&48  		&38,40	\\
Phase center (J2000)&\multicolumn{2}{c}{5$^{\textnormal{h}}$35$^{\textnormal{m}}$18$\rlap{.}^{\textnormal{s}}$30, $-$05$^{\circ}$00$^{\prime}$33$\dotarcsec$01}\\
FWHM of Primary beam (arcsec)  &27		&27\\
PWV (mm)  & $\sim$2.6 	&  1.1$-$2.2   	\\
Phase stability rms (degree)$^a$ &$\sim$13& 21$-$52 	\\
%Polarization calibrator		& J0522-3627	& J0522-3627       &	- 	\\
Bandpass calibrators  &J0522-3627 & J0510+1800\\
Flux calibrator & J0522-3627& J0510+1800, J0522-3627 \\
Phase calibrators$^b$  &J0541-0541 & J0607-0834\\
{Central frequency USB/LSB (GHz)} & \multicolumn{2}{c}{230.535; 231.319 / 217.102; 219.557}\\
Total continuum bandwidth; USB+LSB (GHz)	& 823   &839	\\
Projected baseline ranges (k$\lambda$)   &  8.6$-$238	& 8.9$-$2418 \\
Maximum recoverable size (arcsec)$^c$	& $\sim$9.8	& $\sim$2.0      \\
On-source time (minutes)      	& $\sim$4    &$\sim$16		\\
\hline \noalign {\smallskip}
\end{tabular}
\end{center}
}
\footnotesize $^a${Antenna-based phase differences measured on the bandpass calibrator.}\\
\footnotesize $^b${The phase calibrator was observed every 8 minutes in each execution.}\\
\footnotesize $^c${Our observations were insensitive to emission more extended than this size scale structure at the 10\% of the total flux density (ALMA Cycle 3 Technical Handbook).}\\ %\citep{wilner1994}
\end{table*}

%HAR-baseline: 12m, 14.4m(MMS2)
%LAR-baseline: 11.87m(MMS2)

\begin{table*}[ht!]
{\scriptsize 
\begin{center}
\caption{Summary of the Presented Images}
\label{tab:line statistics}
\begin{tabular}{lcccc}
\hline\hline \noalign {\smallskip}
Image Name& rms Noise Level & Synthesized Beam Size$^a$&Velocity Width&Figure\\
& &(arcsec $\times$ arcsec, degree)& (km\,s$^{-1}$)\\
%&&(blue\&red), (cyan\&magenta)$^b$&&\\
\hline \noalign {\smallskip}
1.3\,mm Continuum &0.61 [mJy\,beam$^{-1}$]&1.52$\times$0.89, $-$78& -& \ref{fig-cont}\,(a)\\
1.1\,mm Continuum &0.04 [mJy\,beam$^{-1}$]&0.14$\times$0.13, $-$54& -& \ref{fig-cont}\,(b)$^{b}$\\
C$^{18}$O Moment 0&22.00 [mJy\,beam$^{-1}$\,km\,s$^{-1}$]&1.59$\times$0.95, $-$78& 0.10& \ref{fig-c18o-mom0}\,(a)$-$(c)\\
C$^{18}$O Moment 1&- &1.59$\times$0.95, $-$78& 0.10& \ref{moment}\,(a)\\
C$^{18}$O Moment 2&- &1.59$\times$0.95, $-$78& 0.10& \ref{moment}\,(b)$-$(c)\\
N$_{2}$D$^{+}$ Moment 0&23.00  [mJy\,beam$^{-1}$\,km\,s$^{-1}$] &1.59$\times$0.95, $-$78& 0.10&\ref{n2d-mom0}\\
C$^{18}$O pv &- &1.59$\times$0.95, $-$78 &-&\ref{n-pv}\\

\hline \noalign {\smallskip}

\end{tabular}
\end{center}
\footnotesize $^a${Briggs weighting (robust = 0.5) was used for the imaging.}\\
\footnotesize $^b${This image is obtained from \cite{liu2024}}.\\
%\footnotesize $^b${Different rms noise levels correspond to the CO integrated intensities obtained by integrating over different velocity ranges in the CO channel maps, which are associated with the outflow and the EHV jet, respectively. }\\
}
\end{table*}

\section{Results}\label{sec:results}

\subsection{1.3\,mm Continuum Emission} \label{result3.1}
%MMS\,2-North is fitted with a deconvolved size of $(0\dotarcsec736\pm0\dotarcsec127)\times(0\dotarcsec179\pm0\dotarcsec089)$ with a P.A. = 108$^{\circ}$$\pm$8$^{\circ}$, which corresponds to $\sim290(\pm50)\times70(\pm35)$\,au in linear size scale. MMS\,2-South is fitted with a deconvolved size of $(0\dotarcsec544\pm0\dotarcsec111)\times(0\dotarcsec305\pm0\dotarcsec053)$ with a P.A. = 102$^{\circ}$$\pm$20$^{\circ}$, which corresponds to $\sim210(\pm44)\times140(\pm21)$\,au in linear size scale.
Figures\,\ref{fig-cont}\,(a) and (b) show the continuum emission obtained from the ALMA 1.3\,mm low angular resolution data (this work) and the 1.1\,mm high angular resolution data \citep{liu2024}, respectively. Figure\,\ref{fig-cont}\,(a) shows the continuum emission toward MMS\,2, which is resolved into two continuum peaks, MMS\,2-North and MMS\,2-South, with a separation of $\sim$1\farcs4 (550\,au). These peak positions and their separation are consistent with those reported for MMS\,2-North-A and MMS\,2-South \citep{liu2024}, where MMS\,2-North corresponds to HOPS-92-A-A and MMS\,2-South corresponds to HOPS-92-B \citep{tobin2020vla}. We performed a two-dimensional (2D) Gaussian fitting using the CASA task ``\texttt{imfit}'' on the 1.3\,mm low angular resolution image (Figure\,\ref{fig-cont}\,(a)) to estimate the sizes of the structures detected in the 1.3\,mm continuum emission associated with MMS\,2-North and MMS\,2-South. The 1.3\,mm continuum fitting results for each source are shown in Table\,\ref{tab:fitting}. The fitted deconvolved sizes of MMS\,2-North and MMS\,2-South are $0.74(\pm0.13)\times0.18(\pm0.09)$ with a position angle (P.A.) $=$ 108$^{\circ}$, and $0.54(\pm0.11)\times0.31(\pm0.05)$ with a P.A. $=$ 102$^{\circ}$, respectively. The corresponding geometric mean deconvolved sizes are 0\farcs36 ($\sim$140 au) for MMS\,2-North and 0\farcs41 ($\sim$160 au) for MMS\,2-South, both of which are smaller than the geometric mean of the synthesized beam of 1\farcs23 ($\sim$480 au) by a factor of $\sim$3. Although the deconvolved size is smaller than the synthesized beam size, \cite{liu2024} suggested that sources with sizes as small as $\sim$1/3 of the synthesized beam can be successfully fitted with uncertainties as small as $\sim$1/70 of the synthesized beam size. The geometric mean sizes of MMS\,2-North and MMS\,2-South derived here are comparable to $\sim$1/3 of the synthesized beam size. We therefore consider the fitting yields a reliable deconvolved size with reasonable uncertainties.

%Since the deconvolved sizes are smaller than the synthesized beam, we therefore adopt the geometric mean of the synthesized beam of 1\farcs23 ($\sim$480\,au) as the upper limit on the diameter of the structures associated with these two sources detected in the 1.3\,mm continuum emission.} 

\begin{table*}[ht!]
{\scriptsize 
\begin{center}
\caption{The 1.3\,mm Continuum Fitting Results}
\label{tab:fitting}
\begin{tabular}{lcccccc}
\hline\hline \noalign {\smallskip}
Source Name& R.A$^{a}$. & Decl$^{a}$. &Total Flux Density& Peak Flux & Deconvolved Size & P.A. \\
& (J2000)& (J2000) & (mJy)&(mJy\,beam$^{-1}$) &(arcsec $\times$ arcsec) & (degree)\\
\hline \noalign {\smallskip}
MMS\,2-North&05 35 18.34& $-$05 00 32.93& $63.5\pm2.5$& $55.9\pm1.2$ & $0.74(\pm0.13)\times0.18(\pm0.09)$$^b$& $108\pm8$\\

MMS\,2-South&05 35 18.27& $-$05 00 32.92& $82.7\pm2.5$& $73.7\pm1.3$ & $0.54(\pm0.11)\times0.31(\pm0.05)$$^c$& $102\pm20$\\
\hline \noalign {\smallskip}

\end{tabular}
\end{center}
\footnotesize $^a${The coordinates refer to the peak position of the source in continuum, which is obtained by the 2D Gaussian fitting.}\\
%\footnotesize $^b${The convolved size shows the upper limit size of the source.}\\
\footnotesize $^{b,c}${The errors of the deconvolved sizes are the 2D Gaussian fitting errors.}\\
\\
}
\end{table*}

Figure\,\ref{fig-cont}\,(b) shows that this binary system could be further resolved into a triple system \citep{tobin2020vla,liu2024}, consisting MMS\,2-North-A, MMS\,2-North-B, and MMS\,2-South \citep{liu2024}. \cite{liu2024} resolved a nearly face-on dust disk associated with MMS\,2-North-A and an inclined dust disk associated with MMS\,2-North-B using the 1.1 mm continuum emission. They also reported that MMS\,2-North-B has a very compact structure, with its disk remaining unresolved.

The continuum peak positions for both MMS\,2-North and MMS\,2-South are consistent between the 1.3\,mm low angular resolution and 1.1\,mm high angular resolution images within the positional accuracy of $\sim$0\dotarcsec005\footnote{The positional accuracy can be approximately estimated as $\sim$(1/2)×($\theta$/(S/N)), where $\theta$ is the synthesized beam size. For our observations, the geometric mean of the synthesized beam is $\theta\sim1\farcs2$.}. MMS\,2-North-B \citep{liu2024} associated with HOPS-92-A-B \citep{tobin2020vla}, is not resolved in the low angular resolution data (this work), likely due to insufficient angular resolution and the faintness of this source. In this paper, we do not further discuss MMS\,2-North-B and refer to the northern component simply as MMS\,2-North. 

%as the low angular resolution image is dominant in MMS\,2-North. 

\subsection{$C^{18}O$(J = 2$-$1) Emission}

\begin{figure*}[ht!]
\vspace{-1\baselineskip}
\gridline{\hspace{-1.5\baselineskip}
          \fig{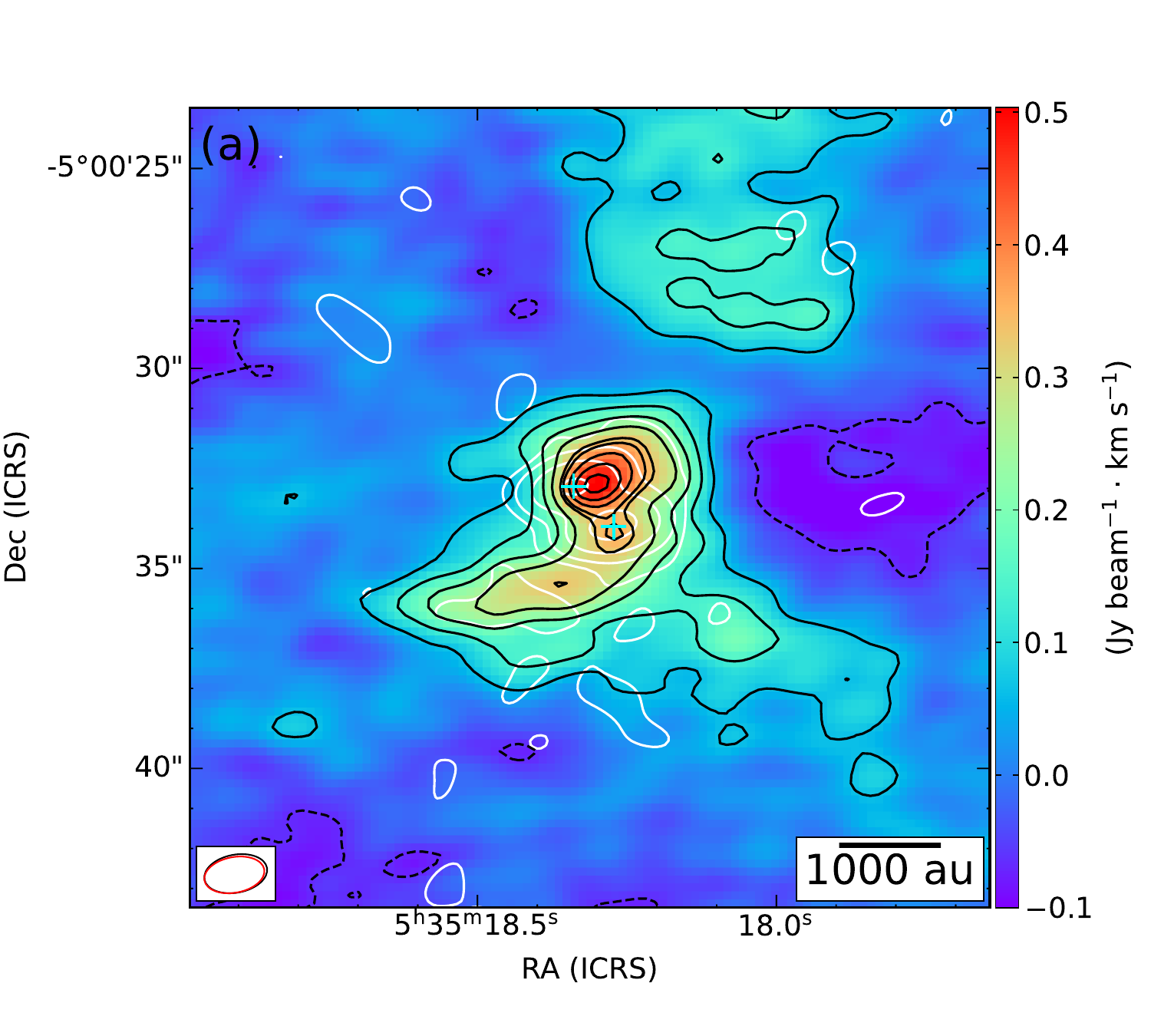}{0.53\textwidth}{}
          \hspace{-0.5\baselineskip}
          \fig{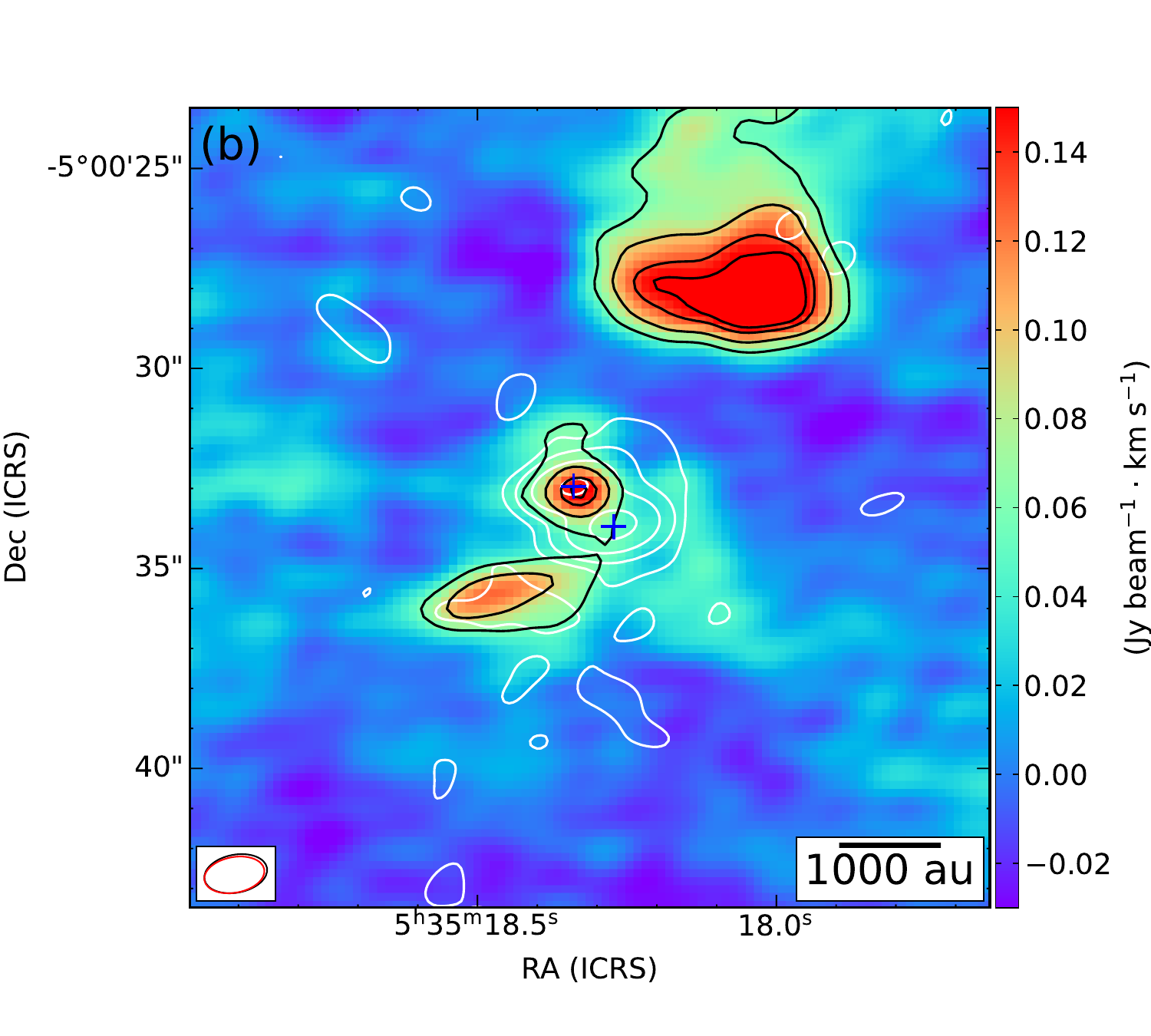}{0.53\textwidth}{}
          }
\vspace{-3\baselineskip}
\gridline{\hspace{-1.5\baselineskip}          
          \fig{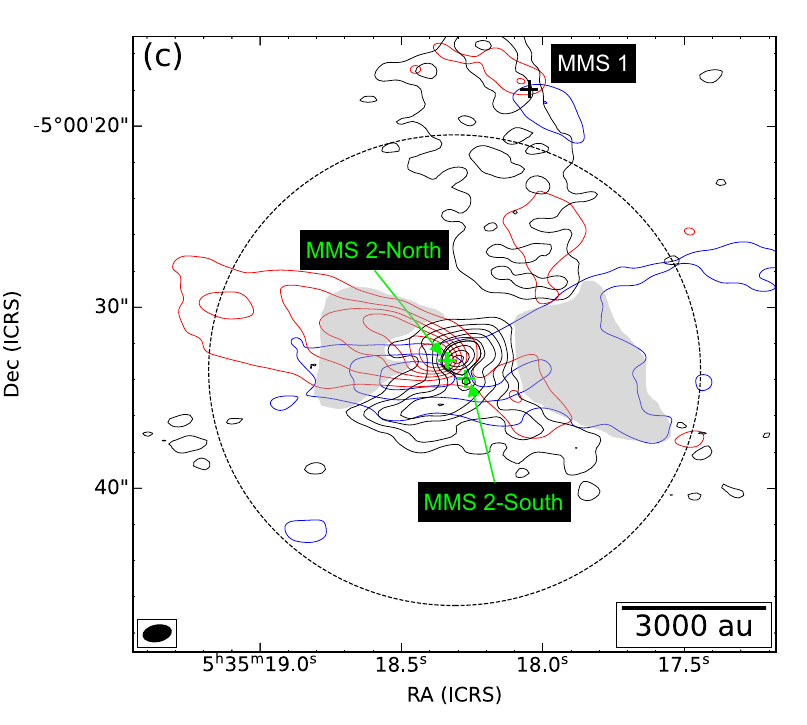}{0.53\textwidth}{}
          }
\vspace{-2\baselineskip}
\caption{(a) Moment 0 (integrated intensity) of C$^{18}$O~($J$ = 2$-$1) in color and black contours overlaid with 1.3\,mm continuum emission in white contours for MMS\,2. The C$^{18}$O emission is integrated from $v_{\textnormal{LRS}}=8.0$ to $12.4$ km\,s$^{-1}$.  The black contour levels are [$-$3, 3, 6, 9, 12, 15, 16, 18, 20, 22] $\times$ $\sigma$ (1$\sigma$ = 22 mJy\,beam$^{-1}$\,km\,s$^{-1}$). The cyan crosses indicate the continuum peak positions associated with MMS\,2-North and MMS\,2-South, respectively. (b) Integrated intensity of C$^{18}$O~($J$ = 2$-$1) in color and black contours overlaid with 1.3\,mm continuum emission in white contours for MMS\,2. The C$^{18}$O emission is integrated from $v_{\textnormal{LRS}}=11.4$ to $11.8$ km\,s$^{-1}$. The black contour levels are [$-$3, 3, 6, 9, 12, 15] $\times$ $\sigma$ (1$\sigma$ = 10 mJy\,beam$^{-1}$\,km\,s$^{-1}$). The blue crosses indicate the continuum peak positions associated with MMS\,2-North and MMS\,2-South, respectively. The white contour levels in panels (a) and (b) are identical to the white contour levels in Figure\,\ref{fig-cont}\,(a). The synthesized beam sizes for C$^{18}$O and 1.3\,mm continuum in panels (a) and (b) are denoted by the open black ellipse and red ellipse in the bottom left corner, respectively. (c) Integrated intensity of C$^{18}$O~($J$ = 2$-$1) in black contours overlaid with integrated intensity of CO~($J$ = 2$-$1) emission in blue and red contours for MMS\,2 \citep{liu2024}. The CO emission is integrated from $v_{\textnormal{LRS}}=-7$ to 10 km\,s$^{-1}$ (blue) and from $v_{\textnormal{LRS}}=12$ to 44 km\,s$^{-1}$ (red), respectively. The blue and red contour levels are [5, 15, 25, 35, 45, 55] $\times$ $\sigma$ (1$\sigma$ = 320 mJy\,beam$^{-1}$\,km\,s$^{-1}$). The black contour levels are [3, 6, 9, 12, 15, 16, 18, 20, 22] $\times$ $\sigma$ (1$\sigma$ = 22 mJy\,beam$^{-1}$\,km\,s$^{-1}$). 
The green crosses indicate the continuum peak positions associated with MMS\,2-North and MMS\,2-South, respectively. The black cross indicates the continuum peak position associated with MMS\,1. The primary beam size for C$^{18}$O is denoted by the circle in black dashed line. The synthesized beam size for both CO and C$^{18}$O is denoted by a filled black ellipse in the bottom left corner. No primary beam correction has been applied to these three images. 
}
\label{fig-c18o-mom0}         
\end{figure*}

\begin{figure*}[ht!]
%\vspace{-25\baselineskip}
\gridline{\hspace{-2\baselineskip}
          \fig{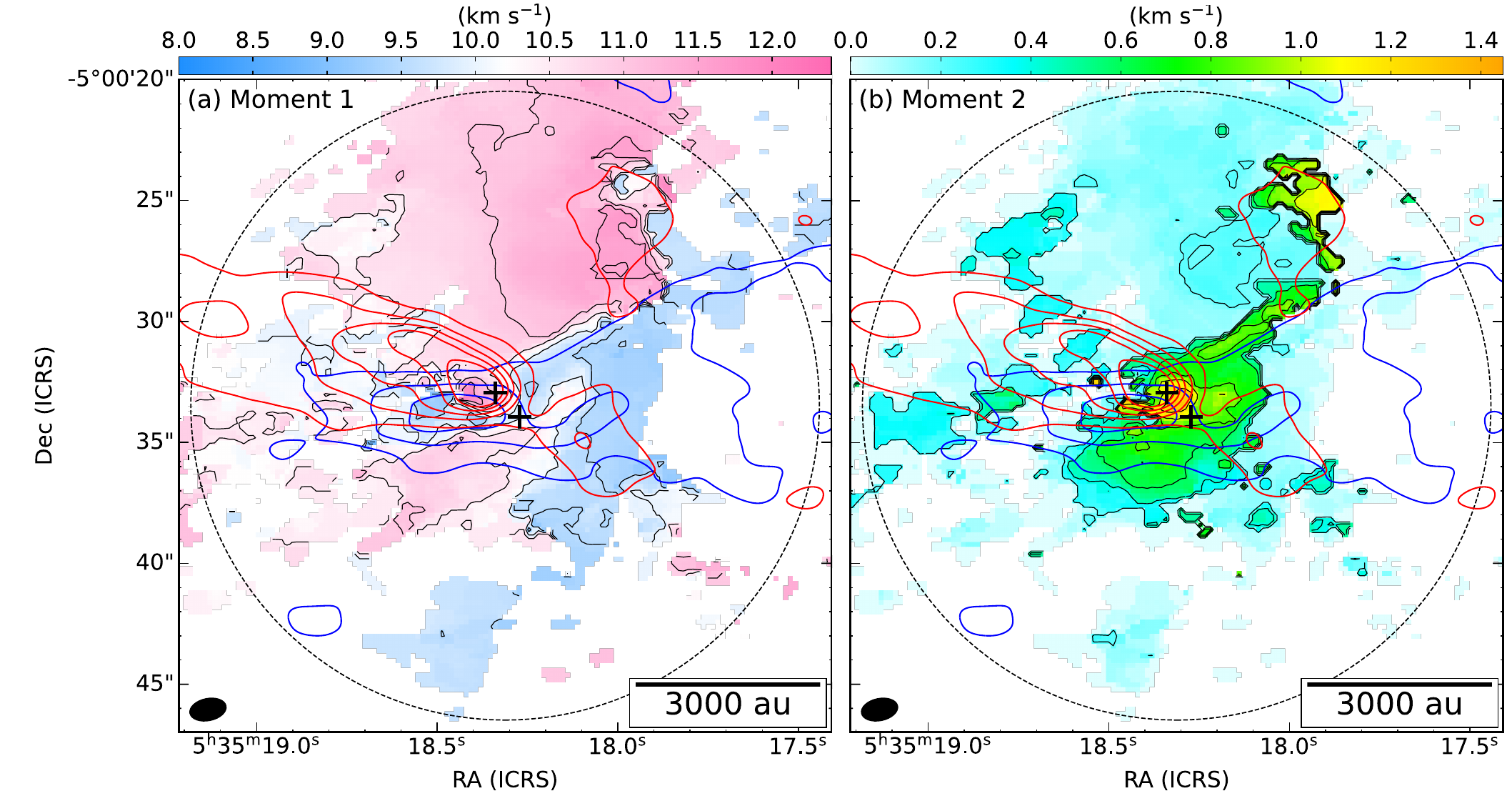}{1.06\textwidth}{}
          }
\vspace{-1\baselineskip}
\caption{Moment 1 (mean velocity) of C$^{18}$O~($J$ = 2$-$1) emission in color and black contours. The black contours are [8.5, 9, 9.5, 10, 10.5, 11, 11.5, 12] $\times$ 1\,km\,s$^{-1}$. The black crosses indicate the continuum peak positions of MMS\,2-North and MMS\,2-South, respectively. (b) Moment 2 (velocity dispersion) of C$^{18}$O~($J$ = 2$-$1) emission in color and black contours. The black contours are [2, 4, 6, 8, 10, 12] $\times$ 0.1\,km\,s$^{-1}$. The cyan crosses indicate the continuum peak positions of MMS\,2-North and MMS\,2-South, respectively. The blue and red contours in panels (a) and (b) are the integrated intensity of blueshifted and redshifted CO~($J$ = 2$-$1) emission, which the contour levels are identical to those in Figure\,\ref{fig-c18o-mom0}\,(c). The primary beam size for C$^{18}$O emission is denoted by the circle in black dashed line in panels (a) and (b). The synthesized beam size for C$^{18}$O emission is denoted by a filled black ellipse in the bottom left corner in all panels. }
\label{moment}          
\end{figure*}

%(a) \textbf{Moment 1 (mean velocity)} of C$^{18}$O~($J$ = 2$-$1) emission in color and black contours. The black contours are [8.5, 9, 9.5, 10, 10.5, 11, 11.5, 12] $\times$ 1\,km\,s$^{-1}$. The black crosses indicate the continuum peak positions of MMS\,2-North and MMS\,2-South, respectively. (b) \textbf{Moment 2 (velocity dispersion)} of C$^{18}$O~($J$ = 2$-$1) emission in color and black contours. The black contours are [2, 4, 6, 8, 10, 12] $\times$ 0.1\,km\,s$^{-1}$. The cyan crosses indicate the continuum peak positions of MMS\,2-North and MMS\,2-South, respectively. (c) Zoomed-in view of panel (b). The color scale of panel (c) is changed from panel (b) for better visualization. The blue crosses indicate the continuum peak positions of MMS\,2-North and MMS\,2-South, respectively. The blue and red contours in panels (a) and (b) are the integrated intensity of blueshifted and redshifted CO~($J$ = 2$-$1) emission, which the contour levels are identical to those in Figure\,\ref{fig-c18o-mom0}\,(c). The primary beam size for C$^{18}$O emission is denoted by the circle in black dashed line in panels (a) and (b). The synthesized beam size for C$^{18}$O emission is denoted by a filled black ellipse in the bottom left corner in all panels. 

Figure\,\ref{fig-c18o-mom0}\,(a) shows the moment 0 (integrated intensity) map of C$^{18}$O emission overlaid with 1.3\,mm continuum emission. The C$^{18}$O emission is integrated over the local standard-of-rest velocities ($v_{\textnormal{LSR}}$) = 8.0$-$12.4\,km\,s$^{-1}$. We detect two centrally condensed structures in C$^{18}$O emission associated with MMS\,2-North, and MMS\,2-South, respectively. The structure associated with MMS\,2-North has a peak intensity detected at an S/N of 22$\sigma$, while the structure associated with MMS\,2-South has a peak intensity detected at an S/N of 16$\sigma$\footnote{The measured positional offsets are therefore larger than the estimated positional uncertainties by factors of $\sim$22 and $\sim$7 for MMS\,2-North and MMS\,2-South, respectively.}. The two C$^{18}$O peaks are separated by $\sim$1\dotarcsec3, comparable to the separation of the two sources in the 1.3\,mm continuum image. The C$^{18}$O intensity peak associated with MMS\,2-North is shifted by $\sim$0\dotarcsec6 to the northwest with respect to the continuum peak, and the C$^{18}$O intensity peak associated with MMS\,2-South is shifted by $\sim$0$\farcs$3 to the south with respect to the continuum peak.

We estimated the sizes of the centrally condensed structures traced by C$^{18}$O emission. For MMS\,2-North, a direct size measurement from Figure\,\ref{fig-c18o-mom0}\,(a) is difficult because the contamination from the extended emission prevents a reliable 2D Gaussian fitting of the centrally condensed structure. Instead, we estimated the size using the channel maps (Figure\,\ref{channel}). Figure\,\ref{channel} shows nearly circular emission on a scale of $\sim$600\,au associated with MMS\,2-North, detected at $\ge$6$\sigma$ with $v_{\textnormal{LSR}}$ = 11.4$-$11.8\,km\,s$^{-1}$. Within this velocity range, the C$^{18}$O intensity peaks are largely consistent with the continuum peak, and the sizes of the structures detected at $\ge$6$\sigma$ are comparable among these velocity channels. To quantify the size of the centrally condensed structure traced by C$^{18}$O, we integrated the C$^{18}$O emission over $v_{\textnormal{LSR}} = 11.4$--$11.8$ km\,s$^{-1}$ as shown in Figure\,\ref{fig-c18o-mom0}\,(b). We estimated the size of the centrally condensed structure associated with MMS\,2-North from Figure\,\ref{fig-c18o-mom0}\,(b) using 2D Gaussian fitting. This structure is fitted with a deconvolved size of 2\farcs11($\pm$0\farcs33) $\times$ 1\farcs27($\pm$0\farcs38) with a P.A. = 45$^{\circ}$, corresponding to a linear size of $\sim$$830 \times 500$\,au. The geometric mean diameter is 1\dotarcsec64 ($\sim$640 au). 

Figure\,\ref{fig-c18o-mom0}\,(a) shows that the diameter of the centrally condensed structure traced by C$^{18}$O associated with MMS\,2-South is unresolved. Therefore, we adopt the beam size of 1\dotarcsec23 ($\sim$480 au), as an upper limit on the diameter of MMS\,2-South.

%\textcolor{red}{Since we estimated the size from limited velocity range, the derived size could be a lower limit.}}

%We directly measured the size of the 6$\sigma$ structure from these channels, and took an averaged value as the diameter $d$. The approximate diameter is $1\dotarcsec7$, which correspond to $\sim$670\,au in linear size. 

In addition, we estimated the size of the extended structure (elongated along the northwest-southeast direction) traced by C$^{18}$O emission using a 2D Gaussian fitting applied to the image shown in Figure\,\ref{fig-c18o-mom0}\,(a). This extended structure likely traces the circumbinary envelope associated with both MMS\,2-North and MMS\,2-South, and is fitted with a deconvolved size of 6\farcs47($\pm$0.56) $\times$ 5\farcs02($\pm$0.45) with a P.A. = 143$^{\circ}$, corresponding to a linear size of $\sim$$2540\times 1970$\,au. The geometric mean diameter is 5\dotarcsec7 ($\sim$2240 au). 

We estimated the physical properties of both centrally condensed structures and the extended structure associated envelope. These properties were derived from the integrated intensity map of the C$^{18}$O emission after applying the primary beam correction. The column density ($N_{\textnormal{H}_{2},\textnormal{C}^{18}\textnormal{O}}$) is derived using the equation obtained from \cite{mangum2015}, which is given by

\begin{equation}
\begin{split}
N_{\textnormal{H}_{2},\textnormal{C}^{18}\textnormal{O}}=
X^{-1}_{\textnormal{C}^{18}\textnormal{O}}\left(\frac{3h}{8\pi^{3}S\mu^2}\right)\times\\
\left(\frac{kT_{\textnormal{ex}}}{hB}+\frac{1}{3}\right)\textnormal{exp}\left(\frac{E_{\textnormal{u}}}{kT_{\textnormal{ex}}}\right)\int T_{\textnormal{B}} dv,
\end{split}
\end{equation}
where $X_{\textnormal{C}^{18}\textnormal{O}}$, $S$, $\mu$, $T_{\textnormal{ex}}$, $E_{\textnormal{u}}$, $T_{\textnormal{B}}$, $h$, $B$, and $k$ are the C$^{18}$O to H$_{2}$ abundance ratio, the line strength, the relevant dipole moment, the excitation  temperature, the energy of upper level above ground, the brightness temperature in units of K, the Planck constant, the rigid rotational constant, and the Boltzmann constant, respectively \citep{mangum2015,feddersen2020carma,morii2021revealing}. 
We adopted $X_{\textnormal{C}^{18}\textnormal{O}}$ = $1.7\times10^{-7}$ \citep{frerking1982},  $S\mu^{2}=0.02$\,Debye$^{2}$, $E_{\textnormal{u}}/k=15.8$\,K, $T_{\textnormal{ex}}=20$\,K, $B=54891.420$\,MHz (obtained from Jet
Propulsion Laboratory (JPL) Molecular Spectroscopy database\footnote{https://spec.jpl.nasa.gov/ftp/pub/catalog/catdir.html}). 

The gas mass ($M_{\textnormal{H}_{2},\textnormal{C}^{18}\textnormal{O}}$) is estimated from the column density,
$N_{\textnormal{H}_{2},\textnormal{C}^{18}\textnormal{O}}$, following \citet{morii2021revealing}, as
\begin{equation}
M_{\textnormal{H}_{2},\textnormal{C}^{18}\textnormal{O}}
= \Omega \mu_{\textnormal{H}_{2}} m_{\textnormal{H}} D^2
N_{\textnormal{H}_{2},\textnormal{C}^{18}\textnormal{O}},
\end{equation}
where $\Omega = \pi \Theta^{2}/(4\ln 2)$ is the solid angle.
The values of $\Theta$ for the centrally condensed structures associated with MMS\,2-North and MMS\,2-South are 1\farcs64 and 1\farcs23, respectively. The values of $\Theta$ for the envelope is 5\farcs70. We adopt a mean molecular weight per hydrogen molecule of $\mu_{\textnormal{H}_{2}} = 2.8$ \citep{kauffmann2008}, a hydrogen atom mass of $m_{\textnormal{H}} = 1.67 \times 10^{-24}$\,g,
and a distance of $D = 393$\,pc \citep{tobin2020vla}.

For MMS\,2-North, we obtain a column density of
$N_{\textnormal{H}_{2},\textnormal{C}^{18}\textnormal{O}} = 1.1 \times 10^{21}$\,cm$^{-2}$
and a gas mass of $M_{\textnormal{H}_{2},\textnormal{C}^{18}\textnormal{O}} = 3.0 \times 10^{-4}\,M_{\odot}$. For MMS\,2-South, the corresponding values are $N_{\textnormal{H}_{2},\textnormal{C}^{18}\textnormal{O}} = 7.0 \times 10^{20}$\,cm$^{-2}$
and $M_{\textnormal{H}_{2},\textnormal{C}^{18}\textnormal{O}} = 1.0 \times 10^{-4}\,M_{\odot}$.
For the extended structure tracing the envelope, estimated using emission above the $3\sigma$ level,
we obtain $N_{\textnormal{H}_{2},\textnormal{C}^{18}\textnormal{O}} = 4.8 \times 10^{21}$\,cm$^{-2}$
and
$M_{\textnormal{H}_{2},\textnormal{C}^{18}\textnormal{O}} = 1.4 \times 10^{-2}\,M_{\odot}$. The diameter $d_{\mathrm{gas}}$, column density
$N_{\textnormal{H}_{2},\textnormal{C}^{18}\textnormal{O}}$,
and gas mass $M_{\textnormal{H}_{2},\textnormal{C}^{18}\textnormal{O}}$ are summarized in Table\,\ref{tab:c18o-properties}.

\begin{table*}[ht!]
{\scriptsize 
\begin{center}
\caption{Physical Properties Derived from C$^{18}$O Emission}
\label{tab:c18o-properties}
\hspace{-3\baselineskip}
\begin{tabular}{lccc}
\hline\hline \noalign {\smallskip}
Name& $d_{\textnormal{gas}}$&$N_{\textnormal{H}_{2},\textnormal{C}^{18}\textnormal{O}}$ & $M_{\textnormal{H}_{2},\textnormal{C}^{18}\textnormal{O}}$$^c$\\
& (au)&($\times10^{21}\textnormal{cm}^{-2}$)& ($\times10^{-3}M_{\odot}$) \\
\hline \noalign {\smallskip}
MMS\,2-North (centrally condensed structure) &640$^a$ & 1.1  &0.3\\
MMS\,2-South (centrally condensed structure) &$\le$480$^b$ & $\le$0.7 & $\le$0.1  \\
MMS\,2 (envelope) & 2240$^c$ &4.8& 14.0\\
\hline \noalign {\smallskip}
\end{tabular}\\
\end{center}
\footnotesize $^a${$d_{\textnormal{gas}}$ is estimated from Figure\,\ref{fig-c18o-mom0}\,(b) using 2D Gaussian fitting.}\\
\footnotesize $^b${$d_{\textnormal{gas}}$ is adopted as the geometric mean of the beam size, providing an upper limit on the source diameter.}\\
\footnotesize $^c${$d_{\textnormal{gas}}$ is estimated from Figure\,\ref{fig-c18o-mom0}\,(a) using 2D Gaussian fitting.}
}
\end{table*}

%\textbf{This offset} and elongation may arise from the contamination by the extended emission, \textbf{which likely traces the gas entrained by the outflow elongated toward northwest} (Figure\,\ref{fig-c18o-mom0}\,(c))

Figure\,\ref{fig-c18o-mom0}\,(c) shows that the extended structure traced by C$^{18}$O emission ($\ge$3$\sigma$ level) is roughly perpendicular to the CO outflow axis \citep[P.A. $\sim70^{\circ}$;][]{liu2024}\footnote{The blueshifted CO emission from MMS\,2-North and MMS\,2-South is merged, making their individual outflow axes difficult to distinguish \citep{liu2024}. We therefore adopt the position angle of $\sim70^\circ$, estimated from the clearly identified redshifted CO emission \citep[$v_{\mathrm{LSR}}-v_{\mathrm{sys}} = 17-31$\,km\,s$^{-1}$;][]{liu2024} associated with MMS\,2-North, as the outflow axis.} and elongated along the northwest--southeast direction, tracing the envelope on a scale of $\sim$2000\,au. The extended structure shows a tail toward the southwest and another toward the southeast. On both the western and eastern sides, the spatial distribution of C$^{18}$O emission is anti‑correlated with the CO outflow, as indicated by the gray‑shaded regions in Figure\,\ref{fig-c18o-mom0}\,(c). Within these regions, C$^{18}$O emission is largely absent. To assess whether this non-detection of C$^{18}$O emission could be attributed to sensitivity limitations, we compared the C$^{18}$O/CO flux ratio over the same velocity range ($v_{\rm LSR}=$ 8–12 km s$^{-1}$). The C$^{18}$O/CO flux ratio toward both the eastern and western sides along the outflow axis is not significantly lower than that in the surrounding regions, suggesting that the non-detection of C$^{18}$O in the outflow component is unlikely to be caused by insufficient sensitivity. This result therefore supports the interpretation that the dense gas traced by C$^{18}$O has likely been swept away by the CO outflow \citep{dobashi1998,arce2006,takahashi2006millimeter}.

We also detect a faint extended structure in C$^{18}$O emission at the 3--6$\sigma$ level, located $\sim$3\arcsec north of the phase center of MMS\,2. The emission is elongated in the north–south direction and is associated with another protostellar system, MMS\,1 \citep{takahashi2024}. However, this emission extends beyond the edge of the primary beam and is disconnected from the emission associated with MMS\,2, so we do not discuss the emission associated with MMS\,1 further in this paper.

%The CO blue-shifted emission is therefore likely is interpreted as entrained gas associated with the CO outflow.

%the local standard-of-rest velocity ($v_{\textnormal{LSR}}$) associated with MMS\,2-North is $\sim$11.0\,km\,s$^{-1}$, while $v_{\textnormal{LSR}}$ of MMS\,2-South is $\sim$9.0\,km\,s$^{-1}$. In addition, there is a clear velocity difference of $\sim$2\,km\,s$^{-1}$ on the northern edge of the blueshifted CO outflow along the north-south direction. 

Figure\,\ref{moment}\,(a) shows the moment 1 (mean velocity) map of the C$^{18}$O emission integrated over $v_{\textnormal{LSR}}$ = 8.0$-$12.4\,km\,s$^{-1}$. The redshifted and blueshifted C$^{18}$O emissions are likely associated with the corresponding redshifted and blueshifted CO emissions, which the CO emissions have the mean velocities $v_{\rm LRS}$ of $\sim$1$-$18\,km\,s$^{-1}$ \citep{liu2024}, larger than the mean velocity of C$^{18}$O emission.

%\textcolor{red}{However, no signatures indicative of envelope rotation, such as a velocity gradient perpendicular to the CO outflow axis, are not clearly observed.}
%The C$^{18}$O emission in the western region \textbf{exhibits} a velocity difference of $\sim$2\,km\,s$^{-1}$ on the side of the blueshifted CO outflow, nearly perpendicular to the CO outflow axis, likely \textbf{resulting} from the interaction with the blueshifted CO outflow. 
%\textbf{Such a velocity structure may arise because, at later evolutionary stages, the dense envelope gas is entrained and swept away by the outflow as the outflow opening angle widens. The remaining dense gas accumulates along the cavity wall, producing a velocity difference perpendicular to the outflow axis \citep{arce2006,takahashi2006millimeter}.} 

Figure\,moment (b) shows the moment 2 (velocity dispersion) map of the C$^{18}$O emission integrated over $v_{\textnormal{LSR}} = 8.0–12.4$\,km\,s$^{-1}$. The central region between the continuum peaks of MMS\,2-North and MMS\,2-South exhibits a relatively high velocity dispersion of $\sim$1.4\,km\,s$^{-1}$. Another component located to the northwest, just outside the blueshifted CO outflow, shows a velocity dispersion of $\sim$1.0\,km\,s$^{-1}$. These two components are likely correlated with the redshifted CO outflow, with the elevated velocity dispersions of $\sim$1.0–1.4\,km\,s$^{-1}$ in the C$^{18}$O emission probably resulting from interaction with the CO outflow. Emission with dispersions of $\sim$0.6–0.8\,km\,s$^{-1}$ is elongated along the northwest–southeast direction, associated with the envelope. Outside the central $\sim$5000\,au region, most of the area shows velocity dispersions of $\sim$0.1–0.3\,km\,s$^{-1}$, consistent with the typical velocity of a cold molecular cloud at a temperature of $\sim$10\,K \citep{Klessen2011}.
%The envelope gas exhibits velocity dispersions of $\gtrsim$0.4\,km\,s$^{-1}$ across a wide area.

%The envelope gas \textbf{exhibits} the velocity dispersions of $\gtrsim$0.4\,km\,s$^{-1}$ \textbf{over a wide area}. The emission with velocity \textbf{dispersions} of $\gtrsim$0.8\,km\,s$^{-1}$ is elongated along the east-west direction and \textbf{tends to align with the CO outflow axis}. \textbf{Outside} the central $\sim$5000\,au region, most of the region shows the velocity \textbf{dispersions} of $\sim$0.1$-$0.3\,km\,s$^{-1}$, consistent with typical velocity of a cold molecular cloud at the temperature of $\sim$10\,K \citep{Klessen2011}. Figure\,\ref{moment}\,(c) further shows that the central region between the continuum peaks of MMS\,2-North and MMS\,2-South \textbf{exhibits} a high velocity dispersion of $\sim$1.4\,km\,s$^{-1}$. The velocity dispersions observed at the continuum peaks \textbf{are $\sim$1.1\,km\,s$^{-1}$ for MMS\,2-North and $\sim$0.9\,km\,s$^{-1}$ for MMS\,2-South, with MMS\,2-North showing slightly higher dispersion.

\subsection{$N_{2}D^{+}$(J = 3$-$2) Emissions}
\begin{figure*}[ht!]
%\vspace{-25\baselineskip}
\gridline{\hspace{-2\baselineskip}
          \fig{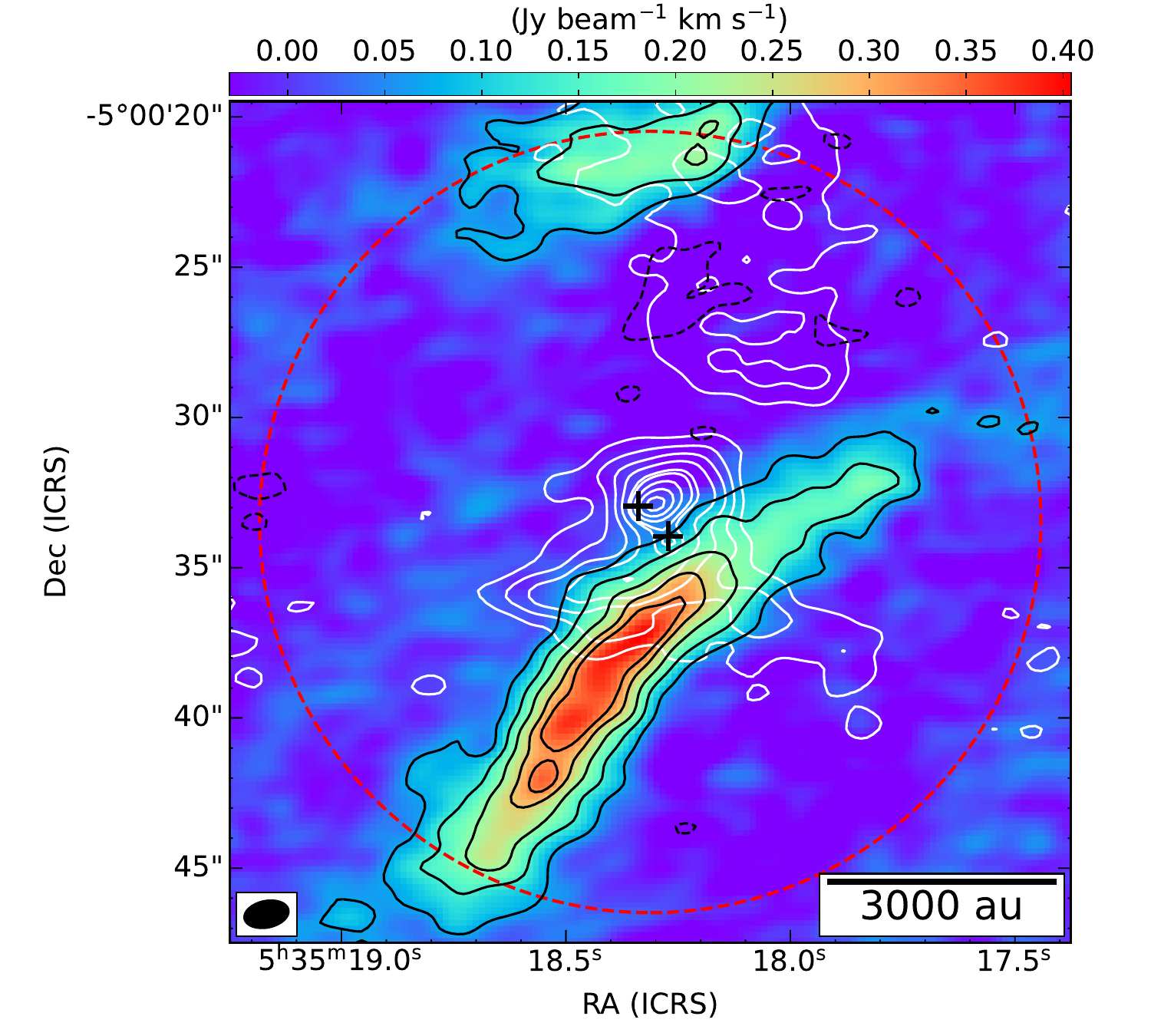}{0.53\textwidth}{}
          }
\vspace{-1\baselineskip}
\caption{
The color and black contours show the moment 0 (integrated intensity) of N$_{2}$D$^{+}$~($J$ = 3$-$2) emission. The white contours are the integrated C$^{18}$O~($J$ = 2$-$1) emission. The black crosses indicate the continuum peak positions of MMS\,2-North and MMS\,2-South, respectively. The black contour levels are [$-$3, 3, 6, 9, 12, 14] $\times$ $\sigma$ (1$\sigma$ = 23 mJy\,beam$^{-1}$\,km\,s$^{-1}$). The white contour levels are [3, 6, 9, 12, 15, 16, 18, 20, 22] $\times$ $\sigma$ (1$\sigma$ = 22 mJy\,beam$^{-1}$\,km\,s$^{-1}$). The primary beam size for N$_{2}$D$^{+}$ is denoted by the circle in red dashed line. The synthesized beam size for N$_{2}$D$^{+}$ emission is denoted by a filled black ellipse in the bottom left corner.
}
\label{n2d-mom0}          
\end{figure*}

Figure\,\ref{n2d-mom0} shows the moment 0 (integrated intensity) map of N$_{2}$D$^{+}$ emission integrated over $v_{\textnormal{LSR}} = 10.3–11.6$\,km\,s$^{-1}$, which is elongated along the northwest-southeast direction with a linear size of $\sim$8600\,au, likely tracing the filamentary structure in the OMC-3 region \citep{chini1997dust,johnstone1999}. Similar structures have also been observed in N$_{2}$D$^{+}$ emission around other millimeter Class 0 sources, including MMS\,3 \citep{morii2021revealing} and MMS\,5 \citep{matsushita2019very}. The N$_{2}$D$^{+}$ emission is bright only on the southern side of MMS\,2-South, while no emission is detected at 3$\sigma$ level near the vicinity of MMS\,2-North. The peak of this elongated structure is offset by $\sim$3\dotarcsec1 to the southeast of the 1.3\,mm continuum peak of MMS 2-South.

\section{Discussion}\label{sec:discussion}

\subsection{Velocity structure of the dense gas traced by C$^{18}$O emission}\label{4.1}

\begin{figure*}[ht!]
%\vspace{-25\baselineskip}
\gridline{\hspace{-2\baselineskip}
          \fig{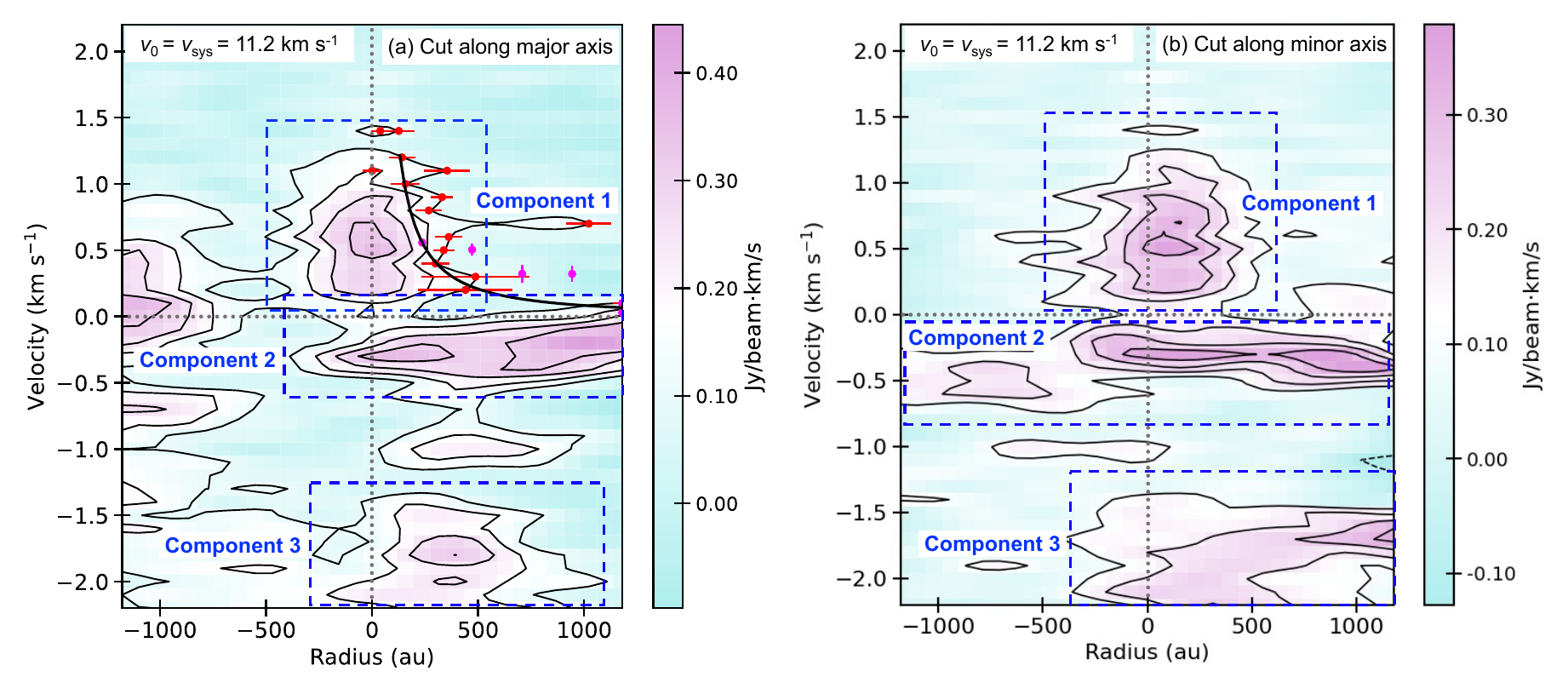}{0.8\textwidth}{}
          }
\vspace{-2\baselineskip}
\caption{
C$^{18}$O\,($J=$2$-$1) PV diagrams along a cut through the 1.3\,mm continuum peak, and oriented along the (a) major axis \citep[P.A. $=149^{\circ}$;][]{liu2024} and, (b) minor axis \citep[P.A. $=59^{\circ}$;][]{liu2024} of dust disk associated with MMS\,2-North, respectively. The black contour levels are [3, 5, 7, 9]$\times$1$\sigma$ (1$\sigma=35$ mJy\,beam$^{-1}$\,km\,s$^{-1}$). The horizontal gray dotted line denotes the systemic velocity ($v_{\textnormal{sys}}$ = 11.2\,km\,s$^{-1}$). The gray vertical dotted line denotes the 1.3\,mm continuum peak position of the MMS\,2-North. The red and magenta dots in panel (a) denote the data points obtained from the SLAM fitting using the ridge and edge methods, respectively. The black curve in panel (a) shows the best-fit model derived using the ridge method. The corresponding model for the edge method is not shown because only a limited number of data points were identified.}
%The blue dashed curve indicates the Keplerian rotationfitting curve ($v_{\phi}\propto r^{-0.5}$). The red dashed curve indicates the conservation of angular momentum fitting line ($v_{\phi}\propto r^{-1}$).}
%Keplerian rotation ($v_{\phi}\propto r^{-0.5}$) and conservation of angular momentum ($v_{\phi}\propto r^{-1}$) assuming the inclination $i$ is 26$^{\circ}$ \citep{liu2024} and the stellar mass $M_{\ast}$ is 0.2 $M_{\odot}$. 
\label{n-pv}          
\end{figure*}

\begin{figure*}[ht!]
\vspace{1\baselineskip}
\gridline{\hspace{-1.5\baselineskip}
          \fig{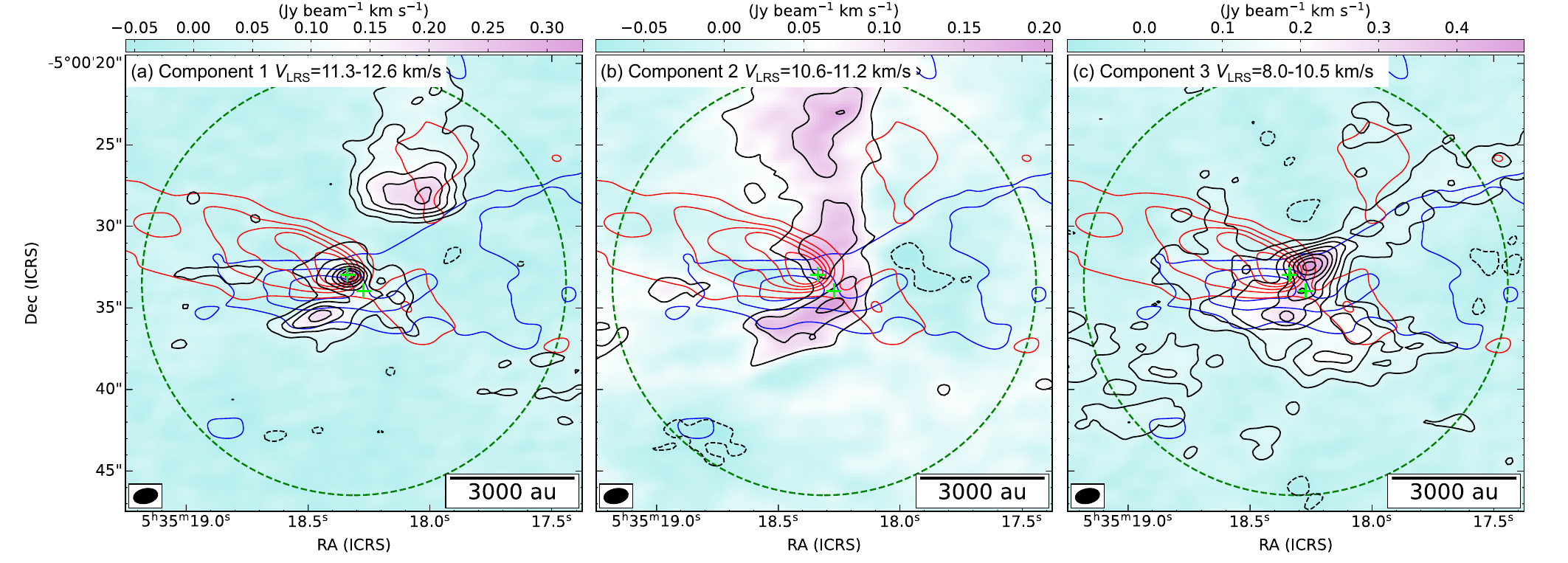}{1\textwidth}{}
          }
\vspace{-2\baselineskip}
\caption{(a) The color and black contours show integrated intensity of Component 1, which is integrated from $V_{\textrm{LRS}} = 11.3-12.6$\,km\,s$^{-1}$. The contour levels are [$-$3, 3, 6, 9, 12, 15, 18, 21] $\times$ $\sigma$ (1$\sigma$ = 20 mJy\,beam$^{-1}$\,km\,s$^{-1}$). (b) The color and black contours show integrated intensity of Component 2, which is integrated from $V_{\textrm{LRS}} = 10.6-11.2$\,km\,s$^{-1}$. The contour levels are [$-$3, 3, 6, 9] $\times$ $\sigma$ (1$\sigma$ = 15 mJy\,beam$^{-1}$\,km\,s$^{-1}$). (c) The color and black contours show integrated intensity of Component 3, which is integrated from $V_{\textrm{LRS}} = 8-10.5$\,km\,s$^{-1}$. The contour levels are [$-$3, 3, 6, 9, 12, 15, 18, 21, 24] $\times$ $\sigma$ (1$\sigma$ = 19 mJy\,beam$^{-1}$\,km\,s$^{-1}$). The blue and red contours in all panels are the blueshifted and redshifted CO emission, which the contour levels are identical to the blue and red contours in Figure\,\ref{fig-c18o-mom0}\,(c). The lime crosses in all panels indicate the peak positions of the continuum emission associated with MMS\,2-North and MMS\,2-South, respectively. The synthesized beam is denoted by an filled black ellipse in the bottom right corner in all panels. The primary beam is denoted by the circle in green dashed line in all panels.}
\label{pv-mom0}
\end{figure*}

To investigate the velocity structure and kinematics of the dense gas structure traced by C$^{18}$O emission, we created position-velocity (PV) diagram using CASA task ``\texttt{impv}''. In this section, we present and discuss only the PV diagram of MMS\,2-North, because the C$^{18}$O emission associated with MMS\,2-South is unresolved, we do not present a PV diagram. Figure~\ref{n-pv} presents the PV diagrams extracted along cuts through the 1.3\,mm continuum peak, oriented along the major axis (P.A. = $149^{\circ}$; \citealt{liu2024}) and minor axis (P.A. = $59^{\circ}$; \citealt{liu2024}) of the dust disk traced by the 1.1\,mm continuum emission, as shown in panels (a) and (b), respectively. The major axis is nearly perpendicular to the redshifted CO outflow axis ($\textrm{P.A.} \sim 70^{\circ}$; \citealt{liu2024}). The PV diagrams reveal three main velocity components: a redshifted component has a velocity range of $v_{\textrm{LRS}}=11.3-12.6$\,km\,s$^{-1}$ (Component\,1), a blueshifted component has a velocity range of $v_{\textrm{LRS}}=10.6-11.2$\,km\,s$^{-1}$ (Component\,2), and an extended blueshifted component has a velocity range of $v_{\textrm{LRS}}=8.0-10.5$\,km\,s$^{-1}$ (Component\,3). The integrated intensity maps of these components are presented in Figure~\ref{pv-mom0}.

%We note that this value lies between 0.5 and 1, suggesting that it reflects a combination of Keplerian rotation ($p\sim0.5$) and angular momentum conservation ($p\sim1$) laws. Because the limited angular resolution, these two components are mixed together. This interpretation also highlights the need for future high angular resolution observations. 
%Both the blueshifted and redshifted sides exhibit intensity peaks. The peak of the redshifted velocity component is located near the continuum peak, whereas the blueshifted component shows a positional offset of $\simeq$500\,au relative to the continuum peak.

Component 1 extends over scales of $\sim$1000\,au and $\sim$1100\,au extracted along the major axis and minor axis, respectively. This component centered around the continuum peak position, which corresponds to the structure condensed around the continuum peak of MMS\,2-North, as shown in Figure\,\ref{pv-mom0}\,(a).

Component 2 extends over scales of $\sim$1500\,au and $\sim$3000\,au extracted along the major axis and minor axis, respectively. Component 2 exhibits a narrowly elongated structure extending through the third and fourth quadrants of the PV diagrams. This component corresponds to the north–south elongated and extended structure shown in Figure\,\ref{pv-mom0}\,(b). As it is detected near the systemic velocity, this component is likely associated with ambient cloud gas.
%Such a non-Keplerian feature could be interpreted as infalling motion toward the central disk \citep[e.g.,][]{yen2014,flores2023}.

Component\,3 extends over scales of $\sim$1200\,au and $\sim$2000\,au extracted along the major axis and minor axis, respectively. Component\,3 shows a blueshifted extended structure on the western side of the continuum peak of MMS\,2-North. Figure\,\ref{pv-mom0}\,(c) further demonstrates that this component is clearly shifted toward the northwest and elongated along the blueshifted CO outflow, suggesting that it is primarily associated with the blueshifted CO outflow. As the outflow gradually clears dense gas over time, the amount of dense gas associated with the envelope decreases, and the remaining material becomes concentrated along the outflow cavity rather than around the protostar. This process is expected to produce an offset between the dense gas and continuum peaks \citep{arce2006}.

Component 1 is centered around the continuum peak, which may be associated with the central protostar. As shown in Figure\,\ref{moment}\,(b), the C$^{18}$O emission exhibits a larger velocity dispersion of $\sim$1.4\,km\,s$^{-1}$ around the continuum peak than in other regions. This enhanced velocity dispersion may be associated with rotational motion of the central protostar. Therefore, we use the public Python package ``\texttt{pvanalysis}'' from the Spectral Line Analysis/Modeling (SLAM) code \citep{slam2023} to fit this component and investigate its kinematic structure and motion. The ``\texttt{pvanalysis}'' firstly identifies the data points from the PV diagram with ridge (peak of the emission) and edge (edge of the emission) methods. These data points are subsequently fitted with a power-law function ($v\propto r^{-p}$). A power-law index of $p\sim 0.5$ indicates the Keplerian rotation, whereas the index of $p\sim 1$ indicates the rotation of infalling material with conserved angular momentum. 

We fitted Component 1 in Figure,\ref{n-pv}(a), extracted along the major axis of the disk, which is nearly perpendicular to the CO outflow axis. We adopted a single power-law model and applied both the ridge and edge methods in ``\texttt{pvanalysis}''. The fit assumes the inclination angle of the dust disk traced by 1.1\,mm continuum $i = 26^{\circ}$ \citep{liu2024}, distance $D = 393$\,pc \citep{tobin2020vla}, and systemic velocity $v_{\textnormal{sys}}$ = 11.2\,km\,s$^{-1}$ \citep{Ikeda2007, tatematsu2008, morii2021revealing}. The ridge method yields a power-law index of $p = 1.3 \pm 0.1$ for this redshifted component. In a Keplerian disk dominated by the gravitational potential of a central protostar, the rotational velocity $v_{\rm rot}$ at radius $r$ around a central star follows $v_{\rm rot}\sim(GM_\ast/r)^{1/2}$, where $G$ is the gravitational constant, $M_\ast$ is the central stellar mass, and the expected rotation profile is $v_{\rm rot}\propto r^{-0.5}$ \citep[e.g.,][]{terebey1984,kido2023,flores2023}. In contrast, for an infalling envelope in which the specific angular momentum $j$ is conserved follows $j\sim rv_{\rm rot}={\rm constant}$, and the expected rotational profile is $v_{\rm rot}\propto r^{-1}$ \citep[e.g.,][]{momose1998}. Our derived slope ($v_{\rm rot}\propto r^{-1.3}$) is more consistent with an infalling material that conserves angular momentum ($v_{\rm rot}\propto r^{-1}$). In addition, the edge method fit was unsuccessful because only a limited number of data points could be identified, as the edges of the PV diagram may be smeared by insufficient angular resolution. We note that, even when applying the ridge method, Keplerian rotation associated with the disk cannot be resolved in the current data. Since typical circumstellar disks have diameters of $\lesssim$100\,au, angular resolutions on comparable or smaller spatial scales are generally required to resolve their Keplerian rotation \citep[e.g.,][]{aso2017,kido2023,flores2023}. Our current observations have a spatial resolution of $\sim620$\,au, which is insufficient to resolve such disk kinematics. Higher angular resolution molecular line observations will therefore be required to confirm the presence of a Keplerian disk in this source.

%The C$^{18}$O intensity peak associated with MMS\,2-North shifts toward northwestern direction by 0\dotarcsec6 with respect to the continuum peak, and the entire structure also shows elongation along this direction. Such an offset and elongation may result from the contamination of the extended emission, as the extended emission is possibly associated with the entrained gas by the outflow elongated to northwestern direction (Figure\,\ref{fig-c18o-mom0}\,(c))

%The redshifted velocity component spreads to $\sim$1.5\,km\,s$^{-1}$ relative to systemic velocity, while the blueshifted one remains close to the systemic velocity with only $\sim$0.5\,km\,s$^{-1}$ deviation. 
%This structure could be a circumatellar disk also include emission from MMS\,2-South, it is difficult to clearly resolve it due to the angular resolution limit. The lower limit of the dynamical mass of the central protostar ($M_{*}$) given by the ridge method is $\sim$0.42\,$M_{\odot}$. This value is about 4 times larger than the protostellar mass estimated for the Class 0 source MMS\,5 \citep{matsushita2019very} in the same region.

%in the inner region up to a breaking radius ($R_b$)The breaking radius ($R_b$) is defined as the radius where the inner and outer power-law indices change (i.e., the power-law index changes from 0.5 (Keplerian rotation) to 1.0 (conservation of angular momentum))} of $\sim$389\,au

\subsection{Comparison of the spatial distribution of N$_2$D$^+$ and C$^{18}$O emissions}
N$_{2}$D$^{+}$ molecule is abundant in the cold ($ T< 20$\,K) and dense ($n=10^{5}$\,cm$^{-3}$) environment \citep[e.g.,][]{emprechtinger} as N$_{2}$D$^{+}$ forms via the ion-molecule reaction between H$_{2}$D$^{+}$ and N$_{2}$ at low temperatures \citep{dalgarno1984,salinas2017}. In such an environment, the molecules CO are frozen onto the dust grains, forming the icy mantles \citep[e.g.,][]{caselli1999}. Therefore, the intense N$_{2}$D$^{+}$ emission is mostly detected around the prestellar sources \citep[e.g.,][]{Crapsi2005,hirano2024}. 

In the protostellar stage, the protostar heats the surrounding region, causing CO to sublimate from dust grains. Gas-phase CO subsequently reacts with and destroys H$_3^{+}$, the parent species of H$_{2}$D$^{+}$, thereby reducing the abundance of H$_{2}$D$^{+}$. Because N$_{2}$D$^{+}$ forms via reactions between H$_{2}$D$^{+}$ and N$_{2}$, its abundance is consequently depleted in warm environments \citep[e.g.,][]{pagani1992,salinas2017}. The offset between the N$_{2}$D$^{+}$ and C$^{18}$O emission peaks around MMS\,2-South could result from the depletion of N$_{2}$D$^{+}$ in warm environments, consistent with previous observations toward the Class 0 source MMS\,3, located next to MMS\,2 in the same filament, where a similar offset between the filamentary structure traced by N$_{2}$D$^{+}$ emission and centrally condensed structure traced by C$^{18}$O emission is detected \citep{morii2021revealing}. However, N$_{2}$D$^{+}$ emission is not detected at $\ge$3$\sigma$ level in the vicinity of MMS\,2-North (e.g., northern side). This is consistent with another Class~0 source located within the same filament, MMS\,5 \citep{matsushita2019very}, where the centrally condensed structure is clearly detected in C$^{18}$O emission but only marginally in N$_{2}$D$^{+}$ emission. The peak of the N$_{2}$D$^{+}$ emission lies near the southern edge of the C$^{18}$O emission. A plausible explanation for the stronger N$_{2}$D$^{+}$ depletion around MMS\,2-North compared to MMS\,2-South, is that the gas near MMS\,2-North is either warmer or exposed to a higher luminosity from its central protostar. Thus, it is expected that increased CO sublimation reduces the abundance of H$_{2}$D$^{+}$ in regions with higher temperatures or higher central stellar luminosities. %\textcolor{blue}{(\textbf{Note:} It is possible that projection effects contribute to the observed offset. However, a more natural and widely accepted explanation is chemical differentiation rather than different locations along the line-of-sight (e.g., Matsushita et al. 2019). As discussed in the manuscript, C18O primarily traces relatively warm ($>$20 K), lower-density gas, whereas N2D+ preferentially forms in cold ($<$20 K), dense regions where CO is depleted through freeze-out.If we were to suggest projection effects as an explanation, we would need supporting kinematic evidence (e.g., multiple velocity components or line-of-sight structures). Given that the referee examine every point carefully, it is very important that any explanation we provide is supported by evidence. Therefore, even if projection effects are only mentioned briefly, we should avoid presenting them as a likely interpretation without additional analysis. For this reason, I wrote "a plausible explanation" rather than implying that it is the primary or only explanation. This wording acknowledges other possible causes (e.g., projection effects) while remaining consistent with the available evidence. Therefore, in this revision, I checked and removed the points or statements that are vague and could not be supported by evidence from this dataset to make this paper more precise and simple.)}

\begin{table*}[ht!]
{\scriptsize 
%\begin{center}
%\raggedleft
\caption{Physical Properties of the Protostellar Source in OMC-3}
\label{tab:source summary}
\hspace{-9\baselineskip}
\begin{tabular}{lccccccccc}

\hline\hline \noalign {\smallskip}
Name& R.A.$^a$  & Decl.$^a$ & Classification$^b$& $L_{\textnormal{bol}}$$^c$ & $M_{\textnormal{H}_2,\textnormal{C}^{18}\textnormal{O}}$$^{e,h}$ &$M_{*}$$^{f,h}$ & $\dot M_{\textnormal{outflow}}$$^{g,h}$ & Jet$^i$ &Reference \\
    &(J2000)& (J2000) & &($L_{\odot}$)&($\times 10^{-3} M_{\odot}$) &$M_{\odot}$&($\times 10^{-6} M_{\odot}\,\textnormal{yr}^{-1}$)&Detection&(case study) \\
    & & & & & & &[Blue, red]& &\\
\hline \noalign {\smallskip}
MMS\,1&5 35 18.03&$-$5 00 17.77 &Class 0 &$<$55$^d$&-&-&-&Yes&\cite{takahashi2024}\\
\hline \noalign {\smallskip}
MMS\,2-North & & & & &0.3 &-&\\
MMS\,2-South &5 35 18.30&$-$5 00 33.01 &Flat-spectrum&17.6&$\le$0.1 &-&[2.5, 5.0] & No&This work\\
MMS\,2 & & & & &14.0 (envelope)&-\\
\hline \noalign {\smallskip}
MMS\,3  & 5 35 18.98&$-$5 00 51.63&Class 0&3.6&890.0 (envelope)&-&-&No&\cite{morii2021revealing}\\
%MMS\,4& 5 35 20.88&$-$5 00 56.25&Prestellar&$<$56$^d$&-&-&-&No&\cite{hirano2024}\\
MMS\,5& 5 35 22.47&$-$5 01 14.43&Class 0&13.8&102.0 (envelope)&0.1&[17.6, 6.0]&Yes&\cite{matsushita2019very}\\
MMS\,6& 5 35 23.42&$-$5 01 30.35&Class 0&31.6&-&-&[12.1, 0.5]&Yes&\cite{takahashi2019alma}\\
MMS\,7& 5 35 26.56&$-$5 03 55.04&Class I&42.9&-&-&[0.4, 14,6]&No&\cite{takahashi2006millimeter}\\

\hline \noalign {\smallskip}

\end{tabular}\\
%\end{center}
\footnotesize $^a${The location of the source denotes the phase center of the ALMA Observations.}\\
\footnotesize $^b${The classification of the protostellar sources are defined based on the SED study presented by \cite{furlan2016herschel}.}\\
\footnotesize $^c${The luminosity $L_{\odot}$ is obtained from \cite{furlan2016herschel}, which can also be obtained from HOPS catalog (https://planetstar formation.iaa.es/OMC-3).}\\
\footnotesize $^d${No counterpart source is identified in the HOPS catalog; hence, $L_{\odot}$ is referenced from \cite{chini1997dust}.}\\
\footnotesize $^e${To avoid the observational bias, only the $M_{\textnormal{H}_2}$ obtained from the data sets with the similar angular resolution are listed.}\\
\footnotesize $^f${The dynamical mass of the central protostar $M_{*}$ is obtained by fitting the PV diagram assuming the Keplerian rotation.}\\
\footnotesize $^g${The mass outflow rate is obtained from Nobeyama 45-m Mapping Observations \citep{tanabe2019nobeyama}.}\\
\footnotesize $^h${“-” means either there is no available values from previous studies.}\\
\footnotesize $^i${Jet refers to the CO emissions detected in the high-velocity range of $v_{\textnormal{LRS}}-v_{\textnormal{sys}}\gtrsim \pm 50$\,km\,s$^{-1}$.}
}
\end{table*}

\subsection{Dense Gas Properties and evolutionary status}
MMS\,2-South is brighter than MMS\,2-North in the continuum, with a slightly higher peak intensity \citep[this work and][]{liu2024}, but it is significantly weaker than MMS\,2-North in C$^{18}$O emission. The estimated gas mass of MMS\,2-South is about three times smaller than that of MMS\,2-North. Such a difference in gas mass may result from the influence of the CO outflow, as much of the dense gas originally associated with MMS\,2-South could have been swept away by the outflow. 

%\textcolor{red}{These differences could result from their different inclinations, as one disk is nearly face-on while the other is inclined, or they could be due to the differences in the projection of the outflows. This hypothesis needs to be confirmed with higher-sensitivity and higher-angular-resolution data in the future}.

%One hypothesis that can be drawn from the current results is that a greater inclination leads to more rapid and extensive sweeping of the surrounding gas.
%There is a discrepancy between the dust disk radius $r_{\textnormal{dust}}$ \citep[see Table~7 in][]{liu2024} and the radius of the gas \textcolor{blue}{disklike} structure, $r_{\textnormal{gas}}$, for MMS\,2-North, where $r_{\textnormal{gas}}$ is estimated as half the value of $d_{\textnormal{gas}}$ listed in Table\,\ref{tab:c18o-properties}. The gas \textcolor{blue}{disklike} structure radius $r_{\textnormal{gas}}$ is 17 times larger than the dust disk radius $r_{\textnormal{dust}}$ for MMS\,2-North \citep[see Table~7 in][]{liu2024}. More evolved sources have shown such discrepancies between dust and gas disks \citep[e.g.,][]{ansdell2018,trapman2020,thieme2023}, because larger grains in the outer dust disk are decoupled from the gas disk and radially drift toward the central protostar, resulting in a reduction in dust disk size \citep[e.g.,][]{weidenschilling1977,birnstiel2010,trapman2020}.This suggests that more evolved sources tend to exhibit a larger discrepancy between their dust and gas disks.

We find that the gas-to-dust ratio derived from $M_{\textnormal{H}_{2},\textnormal{C}^{18}\textnormal{O}}$ and $M_{\textnormal{dust,1.3\,mm}}$ is $\sim$1.4 for MMS\,2-North, whereas MMS\,2-South exhibits an even lower value of $\sim$0.3. Here, $M_{\textnormal{dust,1.3\,mm}}$ is derived from the 1.3\,mm continuum emission in Figure\,\ref{fig-cont}\,(a). Although the 1.3\,mm continuum data do not have sufficient angular resolution to resolve the dust disk, the estimated values still reflect the local gas-to-dust ratios on the beam scale ($\sim$600\,au) around MMS2-North and MMS2-South. Both ratios are significantly lower than the typical ISM gas-to-dust ratio of $\sim$100. Similarly low gas-to-dust ratios have been reported for disks in the Lupus star-forming region, where gas masses were derived from $^{13}$CO and C$^{18}$O emissions. In that sample, gas-to-dust ratios span down to $\sim$0.1, with most disks exhibiting values of $\sim$1--10 \citep{miotello2017}. One possible explanation for the low gas-to-dust ratio is that optically thin CO isotopologues may underestimate the disk gas mass due to gas dissipation or carbon depletion \citep[e.g.,][]{williams2014,miotello2017}. We caution that the derived gas-to-dust ratio, is based on the assumption that the C$^{18}$O emission traces the gas disk and may therefore be affected by uncertainties arising from the limited angular resolution of the present observations. Future studies using higher angular resolution molecular line observations and additional gas tracers with chemical properties similar to H$_2$, such as hydrogen deuteride \citep[HD;][]{trapman2017}, will help to better constrain the gas-to-dust ratio.

To minimize the observational bias, we compare the dense gas properties of MMS\,2 with those of the younger sources that were observed within the same project. The relevant source properties are summarized in Table\,\ref{tab:source summary}. The envelope traced by C$^{18}$O emission associated with MMS\,2 has a diameter smaller by factors of $\sim$5 and $\sim$2, compared to those of the Class~0 sources MMS\,3 \citep{morii2021revealing} and MMS\,5 \citep{matsushita2019very} in the OMC-3 region, respectively. The C$^{18}$O gas mass of the envelope associated with MMS\,2 is also lower by factors of $\sim$60 and $\sim$7, compared to those of the MMS\,3 \citep{morii2021revealing} and MMS\,5 \citep{matsushita2019very}, respectively. These differences suggest that the more dense envelope gas in MMS\,2 has been dissipated. We have also identified the interaction between the dense gas and the outflow in MMS\,2-North, as discussed in Sec.\,\ref{4.1}. The C$^{18}$O emission associated with MMS\,2-North is clearly shifted toward the northwest and elongated along the blueshifted CO outflow. Because MMS\,2, MMS\,3, and MMS\,5 were observed as part of the same project with the same angular resolution, the differences in envelope size and mass are unlikely to be caused by observational biases. Moreover, MMS\,2 is at a later evolutionary stage than MMS\,3 and MMS\,5. Therefore, these results suggest that the envelope of MMS\,2 contains less dense gas, likely as a result of envelope dissipation and clearing by the CO outflow during its later evolutionary stage.

%the C$^{18}$O emission peaks associated with MMS\,2-North and MMS\,2-South are clearly offset from their continuum peaks, suggesting that the dense envelope gas has likely been swept \textbf{away} by the outflow. As the outflow gradually clears dense gas over time, the amount of dense gas associated with the envelope decreases, and \textbf{the remaining material becomes concentrated along the outflow cavity} rather than around the protostar. This process is expected to produce an offset between the dense gas and continuum peaks \citep{arce2006}

%Combined with the detection of a CO outflow toward MMS\,2, this may indicate that material continues to accrete toward the central region, causing the central protostar and disk to grow in both size and mass even in the late evolutionary phase. This interpretation is consistent with previous suggestion that the size and mass of dust disks evolves with protostellar evolutionary stages \citep[e.g.,][]{yen2017,liu2024}. %These results suggest that gas disk is generally larger than dust disk and likely to continue to accrete in the late flat-spectrum phase.

The detection of CO outflow toward MMS\,2 suggests that material may still be accreting onto the central region even at the late evolutionary phase. However, no high velocity jet, such as the one detected in CO and SiO emission toward MMS\,5 \citep{matsushita2019very}, was found in MMS\,2. Moreover, the averaged mass outflow rate ($\dot M_{\textnormal{outflow}}$) of the blueshifted and redshifted lobes is lower by a factor of $\sim$2$-$3 compared with values measured for Class 0/I sources in the same region \citep[e.g.,][]{takahashi2008millimeter,tanabe2019nobeyama}. The corresponding $\dot M_{\textnormal{outflow}}$ values are listed in Table\,\ref{tab:source summary}. This indicates that the outflow/jet activity in MMS\,2 is less prominent than those observed in Class~0/I sources in this region, such as MMS\,1 \citep{takahashi2024}, MMS\,5 \citep{matsushita2019very}, and MMS\,6 \citep{takahashi2012molecular}. This suggests that MMS\,2 may have a lower accretion rate than the younger Class 0/I sources, as the outflows in YSOs is considered to be regulated by mass accretion process \citep{calvet2004}.

% The dynamical mass (assuming the Keplerian rotation) of the central protostar of MMS\,2-North is more massive than the Class 0 source MMS\,5 by a factor of $\sim$4 \citep{matsushita2019very}.

%As shown in Figure\,\ref{mdot-co}, the mass outflow rate $\dot M$ tentatively decreases with the evolutionary stage. 

%remove mms6

%are close to those of the T Tauri stars TW\,Hya, DM\,Tau, and GM\,Aurit as these sources show the N$_2$H$^{+}$/C$^{18}$O gas masses within the range of $\sim$$10^{-1}-10^{-3}$ \citep{Trapman2022}, but they 
%Our class I/flat source has also shown the flatten structure, which was mostly seen in the more evolved class II stage. More evolved source (e.g.,Class II) shows the flatten circumstellar structures, as most of the dense gas was swept away by the outflow, only the dense gas associated with the disk were left shows structure perpendicular to outflow (Reference). Our class I/flat source has also shown the flatten structure.

%\begin{figure*}[ht!]
%\vspace{-25\baselineskip}
%\gridline{\hspace{-2\baselineskip}
%          \fig{Mdot-CO.pdf}{0.53\textwidth}{}
%          }
%\vspace{-1\baselineskip}
%\label{mdot-co}          
%\end{figure*}

\section{Conclusions}\label{sec:conclusion}
We present the ALMA 1.3\,mm continuum and C$^{18}$O ($J=2-1$) and N$_2$D$^+$ ($J=3-2$) observations with an angular resolution of $\sim$1\dotarcsec59 (620 au) toward MMS\,2, a millimeter flat-spectrum multiple system in the OMC-3 region. The results are summarized as follows.
\begin{enumerate}
    \item We detect 1.3\,mm continuum emission toward MMS\,2-North and MMS\,2-South. The fitted geometric mean deconvolved sizes are 0\farcs36 ($\sim$140\,au) for MMS\,2-North and 0\farcs41 ($\sim$160\,au) for MMS\,2-South, both of which are smaller than the geometric mean of the synthesized beam of 1\farcs23 ($\sim$480\,au) by a factor of $\sim$3. Their intensity peak positions align with previous 1.1\,mm high angular resolution observations \citep{liu2024}. The extremely compact and faint source MMS\,2-North-B reported by \cite{liu2024} could not be resolved in this work, due to the angular resolution limits.
    
    \item We detect the centrally condensed structures associated with MMS\,2-North and MMS\,2-South in the C$^{18}$O emission. The centrally condensed structure around MMS\,2-North has an estimated diameter of $\sim$1\dotarcsec64 ($\sim$640\,au), and a corresponding mass of $3.0 \times 10^{-4}\,M_{\odot}$, respectively. For MMS\,2-South, both the diameter and mass of the centrally condensed structure remain unresolved, with upper limits of $\sim$1\dotarcsec23 ($\sim$480\,au) and $1.0 \times 10^{-4}\,M_{\odot}$, respectively. 
    %\textcolor{blue}{(Note: I have removed the phrase "centrally condensed structure" when describing the 1.3 mm continuum and now use this term only for the small-scale C18O emission. I also avoided using a phrase such as "We have detected C18O emission toward MMS2-North with the extension of 600 au." because the following sentence already provides a more precise estimate of the structure's size (e.g., 1.64"($\sim640$au)) and mass, making the additional description redundant.)}
    %\textcolor{red}{The current data do not resolve the Keplerian rotation associated with the circumstellar disk because of the limited angular resolution}.
    %The gas masses derived from C$^{18}$O may underestimate the actual disk masses due to gas dissipation \textbf{or} carbon depletion.

    \item We detect the extended structure associated with both MMS\,2-North and MMS\,2-South in the C$^{18}$O emission, likely tracing the circumbinary envelope. The geometric mean diameter of this structure is 5\dotarcsec70 ($\sim$2240)\,au, with an estimated gas mass of $1.4 \times 10^{-2}$\,$M_{\odot}$. This extended structure is oriented roughly perpendicular to the CO outflow. The PV diagram fitting result suggests the kinematics likely consistent with a rotating infalling envelope.

    %\textcolor{red}{No clear signatures indicative of envelope rotation, such as a velocity gradient perpendicular to the CO outflow axis, are observed.}
    %\textcolor{red}{This extended C$^{18}$O structure may have been dissipated or swept up by the outflows}.
 
    \item We detect a northwest-southeast filamentary structure in N$_{2}$D$^{+}$ emission, with a length of $\sim$8600\,au in linear size scale on the southern side of MMS\,2-South. The offset between the N$_{2}$D$^{+}$ and C$^{18}$O emissions likely results from the warm protostellar environments. Additionally, the stronger depletion of N$_{2}$D$^{+}$ toward MMS\,2-North relative to MMS\,2-South may be attributed to the higher luminosity and higher temperature of MMS\,2-North.
 
    \item The flat-spectrum source MMS\,2 likely possesses less dense envelope gas than the Class~0/I sources observed within same region and in the same project, suggesting that a larger amount of dense envelope gas has been dissipated and swept up by the CO outflow during this later evolutionary stage. Moreover, material may continue to accrete onto the central region of MMS\,2 in the flat-spectrum phase. However, the accretion rate is likely lower than that of younger Class~0/I sources, as suggested by the lower mass outflow rate and non-detection of jet toward MMS\,2.

%The gas-to-dust ratio derived in this study is significantly lower than the canonical ISM value. Whether this reflects gas dissipation or carbon depletion can be tested with \textbf{higher angular resolution and higher sensitivity observations} using alternative gas tracers in the future. 

%Last but not least, the large differences in gas properties and kinematics between MMS\,2-North and MMS\,2-South suggest the inclinations could significantly affect the dense gas structures in the late stage. This might also impact on the initial conditions of planet formation, which should also be investigate in the future.
    
\end{enumerate}

\section*{Acknowledgments}
We thank the referee for the thorough review and for providing the helpful and detailed comments. This paper makes use of the following ALMA data: ADS/JAO.ALMA\#2015.1.00341.S. ALMA is a partnership of ESO (representing its member states), NSF (USA), and NINS (Japan), together with NRC (Canada), MOST and ASIAA (Taiwan), and KASI (Republic of Korea), in cooperation with the Republic of Chile. The Joint ALMA Observatory is operated by ESO, AUI/NRAO, and NAOJ. This work has been supported by the Strategic Priority Research Program of the Chinese Academy of Sciences (CAS) Grant No.\ XDB0800300, the National Natural Science Foundation of China (NSFC) through grant Nos.\ 12273090 and 12322305, and the Natural Science Foundation of Shanghai (No.\ 23ZR1482100). Data analysis was carried out on the Multi-wavelength Data Analysis System operated by the Astronomy Data Center (ADC), National Astronomical Observatory of Japan. This work was supported by a NAOJ ALMA Scientific Research grant (No.2022-22B).\\
\facilities{ALMA}
\software{astropy \citep{astropy2013}, APLpy \citep{robitaille2012}, CASA \citep{bean2022casa}, matplotlib \citep{hunter2007}, SLAM\citep{slam2023}}

\bibliography{reference}
\bibliographystyle{aasjournalv7}

\clearpage

\appendix
\section{C$^{18}$O Channel Maps}\label{apdx-A}
\renewcommand{\thefigure}{A\arabic{figure}} 
\renewcommand{\theHfigure}{A\arabic{figure}}
\setcounter{figure}{0}
We present the C$^{18}$O channel maps for MMS\,2 in this section. The velocity width is 0.1\,km\,s$^{-1}$.

\begin{figure*}[ht!]
%\vspace{-25\baselineskip}
\gridline{\hspace{-1.5\baselineskip}
          \fig{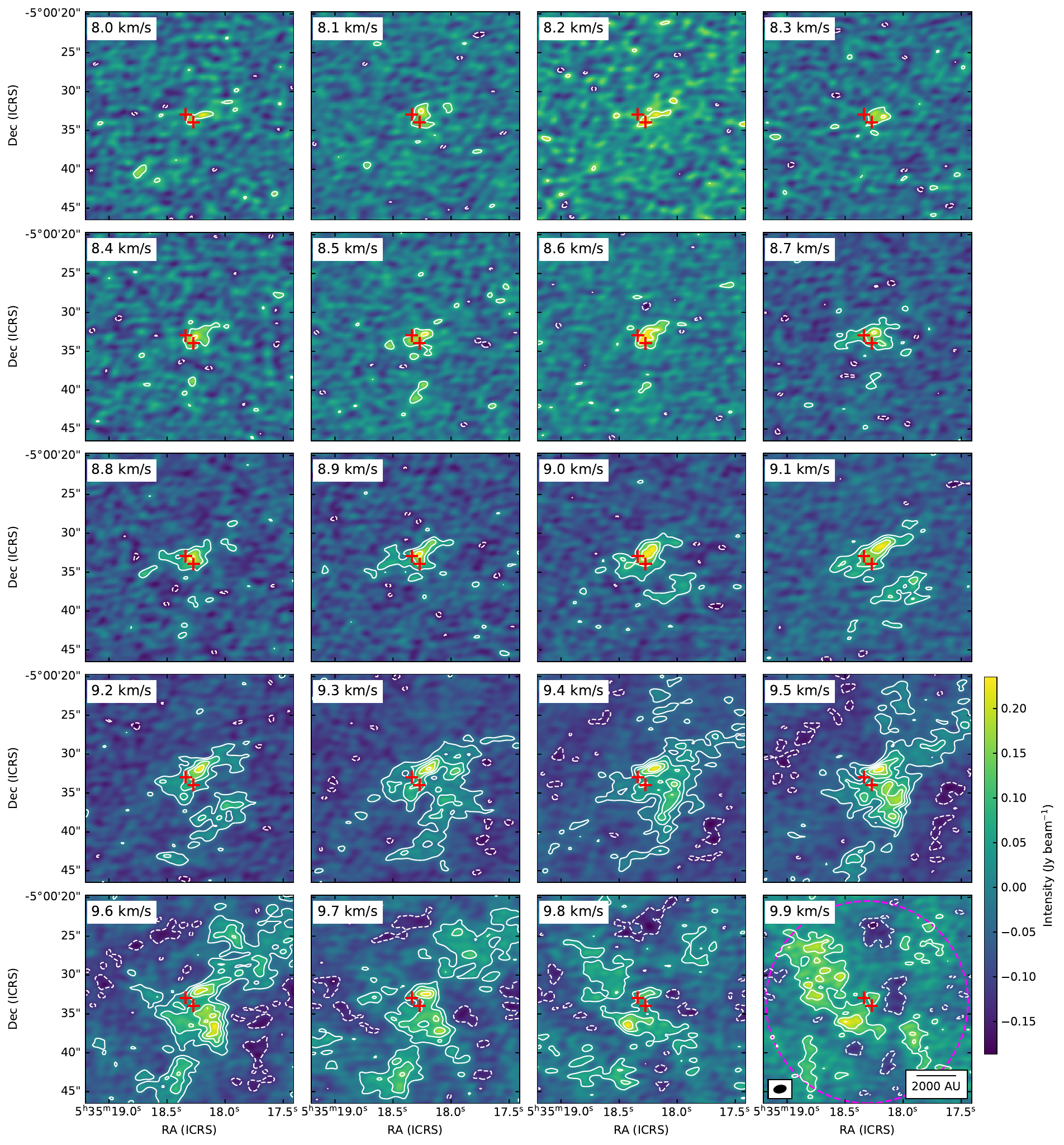}{0.92\textwidth}{}
          }
\vspace{-2\baselineskip}
\caption{Channel maps of C$^{18}$O~($J$ = 2$-$1) emission shown in color and white contours. The $v_{\textnormal{LSR}}$ is shown in the upper left corner. The contour levels are [-6, -3, 3, 6, 9, 12, 15] $\times$ $\sigma$ (1$\sigma$ = 0.026 Jy\,beam$^{-1}$). The red crosses indicate the peak positions of the continuum emission associated with MMS\,2-North and MMS\,2-South, respectively. The synthesized beam is denoted by an filled black ellipse in the bottom right corner. The primary beam is denoted by the circle in magenta dashed line. }
\label{channel}
\end{figure*}

\setcounter{figure}{0}
\begin{figure*}[ht!]
\vspace{4\baselineskip}
%\vspace{-25\baselineskip}
\gridline{\hspace{-1.5\baselineskip}
          \fig{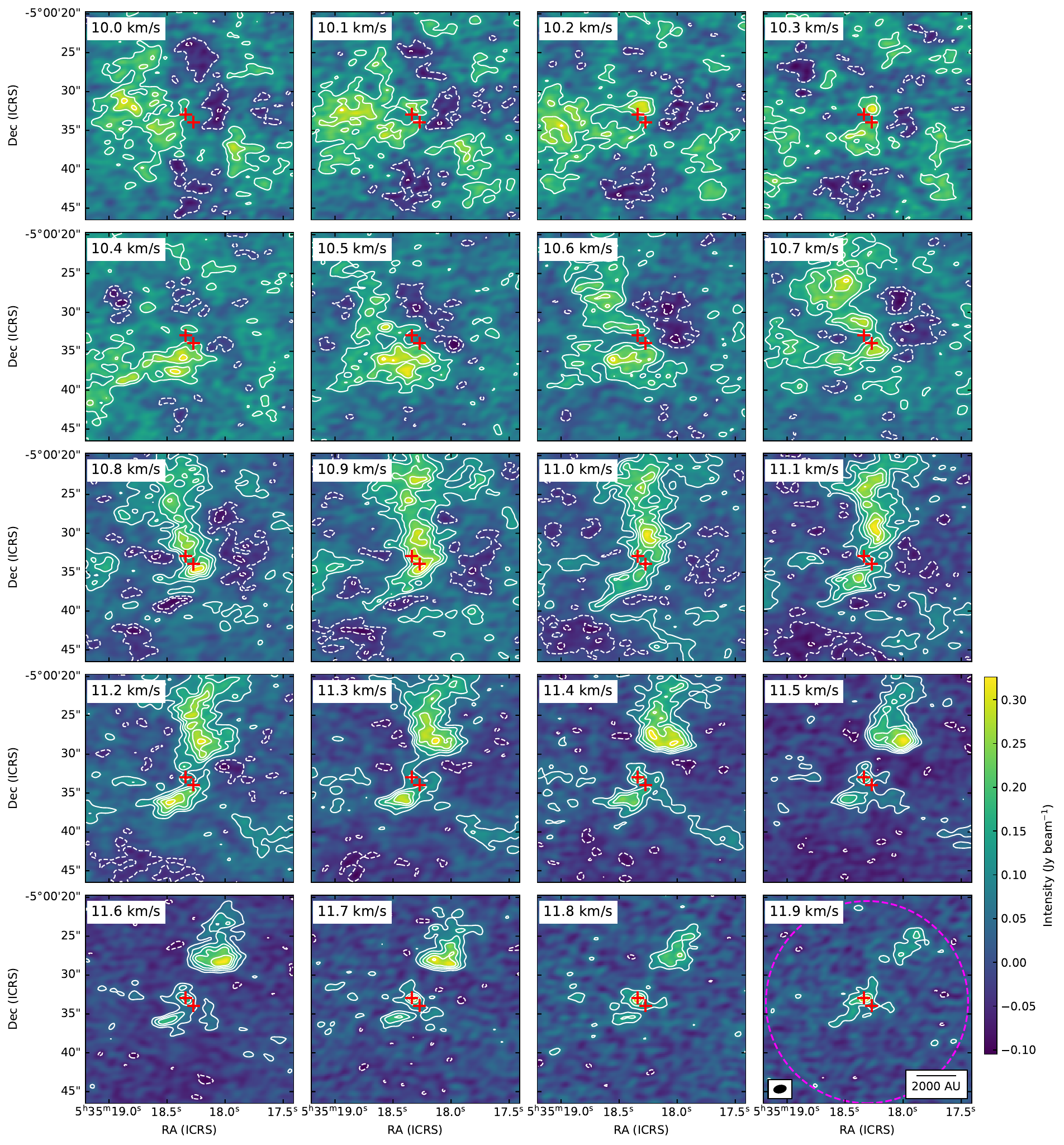}{0.92\textwidth}{}
          }
\vspace{-1\baselineskip}
\caption{(continued)}
\end{figure*}
\vspace{2\baselineskip}

\setcounter{figure}{0}
\begin{figure*}[ht!]
%\vspace{-25\baselineskip}
\gridline{\hspace{-1.5\baselineskip}
          \fig{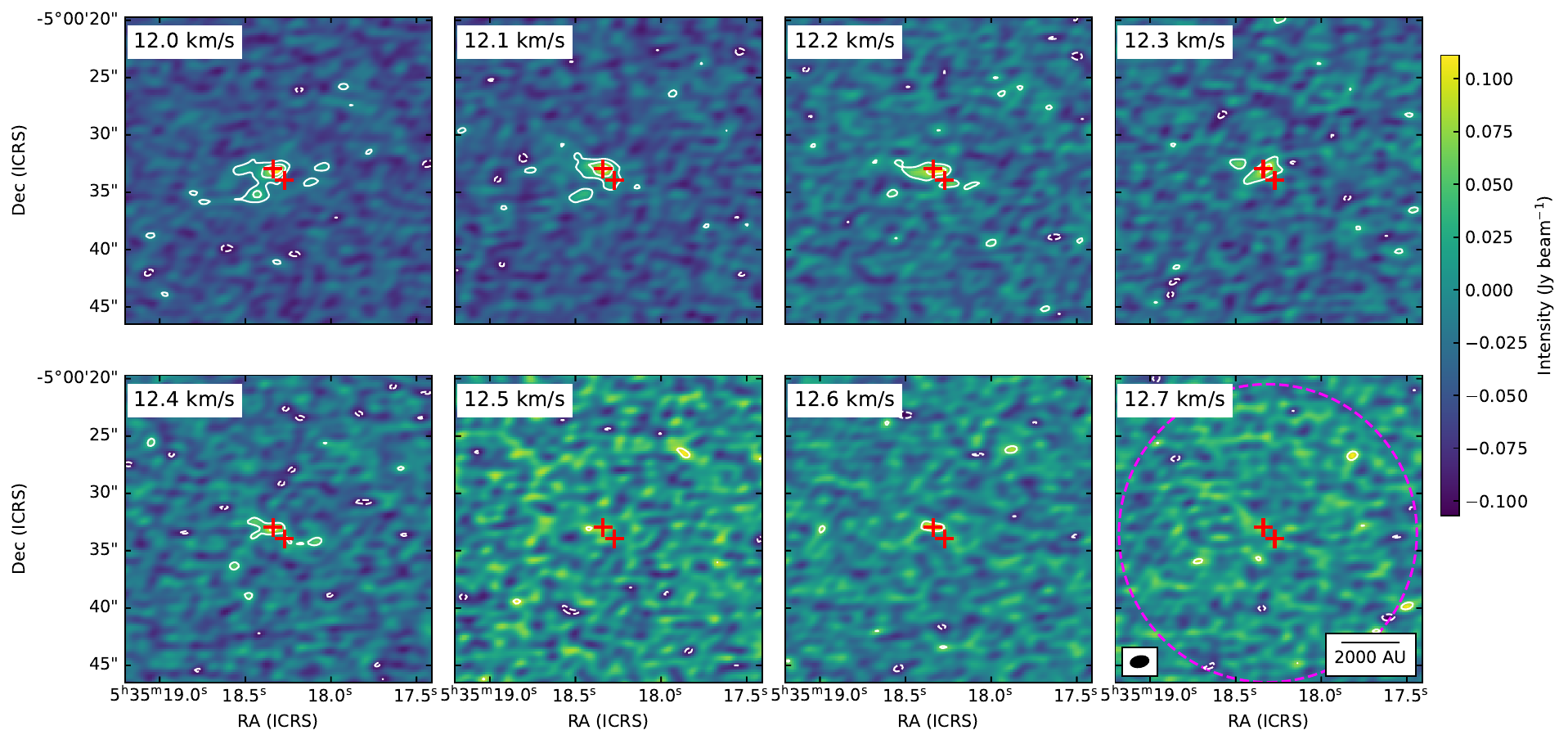}{1\textwidth}{}
          }
\vspace{-1\baselineskip}
\caption{(continued)}
\end{figure*}

%\section{Integrated Intensity Maps of PV Components}\label{apdx-B}
%\renewcommand{\thefigure}{B\arabic{figure}} 
%\renewcommand{\theHfigure}{B\arabic{figure}}
%\setcounter{figure}{0}

%\textbf{We present the integrated intensity maps of the Component 1, Component 2, and Component 3 identified from C$^{18}$O PV diagram (Figure\,\ref{n-pv}), respectively}.

\end{document}